\documentclass[reprint,preprintnumbers,nofootinbib,amsmath,amssymb,floatfix,aps,longbibliography,prd]{revtex4-1}
\usepackage{graphicx}
\usepackage{subfigure}%
\usepackage{dcolumn}
\usepackage{isotope}
\usepackage{multirow}
\usepackage{textcomp}
\usepackage{gensymb}
\usepackage{xcolor}
\usepackage{hyperref}

\newcommand{\etal}{{\it et al.}}
\newcommand{\GENIE}{\textsc{genie}}
\newcommand{\genie}{\textsc{genie}}
\newcommand{\nova}{{NOvA}}

\newcommand{\pythia}{\textsc{pythia}}
\newcommand{\ve}[1]{\ensuremath{\mathbf{#1}}}
\newcommand{\n}[1]{\ensuremath{|\mathbf{#1}|}}

\begin{document}

\title{Assessing the accuracy of the GENIE event generator with electron-scattering data}
\author{Artur M. Ankowski}
\author{Alexander Friedland}
\affiliation{SLAC National Accelerator Laboratory, Stanford University, Menlo Park, California 94025, USA}
\date{July 30, 2020}

\preprint{SLAC-PUB-17541}

\begin{abstract}
Precision neutrino oscillation experiments of the future---of which DUNE is a prime example---require reliable event generator tools. The 1--4 GeV energy regime, in which DUNE will operate, is marked by the transition from the low-energy nuclear physics domain to that of perturbative QCD, resulting in rich and highly complex physics. Given this complexity, it is important to establish a validation procedure capable of disentangling the physical processes and testing each of them individually. Here, we demonstrate the utility of this approach by benchmarking the GENIE generator, currently used by all Fermilab-based experiments, against a broad set of inclusive electron-scattering data. This comparison takes advantage of the fact that, while electron-nucleus and neutrino-nucleus processes share a lot of common physics, electron scattering gives one access to precisely known beam energies and scattering kinematics. Exploring the kinematic parameter range relevant to DUNE in this manner, we observe patterns of large discrepancies between the generator and data. These discrepancies are most prominent in the pion-producing regimes and are present not only in medium-sized nuclei, including argon, but also in deuterium and hydrogen targets, indicating mismodeled hadronic physics. Several directions for possible improvement are discussed.
\end{abstract}

\maketitle

\section{Introduction}

Recent years have seen a resurgence of interest in the physics of neutrino-nucleus interactions.
The motivation comes from the wealth of data generated by modern neutrino experiments, such as T2K, MiniBooNE, MicroBooNE, NOvA, MINERvA, and the development of the DUNE and Hyper-Kamiokande research programs. NOvA, MINERvA, and DUNE, in particular, operate in neutrino beams of a few GeV energy, where the physics of neutrino scattering is especially involved.
Whether one thinks of this regime as ``high-energy nuclear physics'' or as ``low-energy QCD'', a first-principles, rigorous description of the full physics is not available.

At the fundamental level, a number of hadronic processes contribute: quasielastic (QE) scattering, resonant and nonresonant pion production, as well as a transition from higher resonances to deep-inelastic scattering (DIS). The problem is further compounded by the complexity of the target nuclei, such as carbon and, even more so, argon. Effects of nuclear Fermi motion, binding energy, correlated nucleon pairs, two-nucleon currents, and final-state interactions can all be important.
Given this complexity of physics, event generator codes, by necessity, use approximate models in each scattering regime and interpolate between them.
It is crucial to be able to assess the accuracy of this treatment.

This task is more complicated than simply comparing generator predictions to neutrino data, for two reasons.
First, generators are often tuned to the same data, making them not usable as validation tools. Second, neutrino scattering measurements are typically reported integrated over a range of energies and angles.
Both of these factors, the tuning and the integrated kinematics, may mask significant problems, which may reappear on different targets or in different kinematic regimes in future analyses. For example, a code tuned to the MINERvA data may not agree with the NOvA or T2K near detector measurements, or the code tuned to  inclusive measurements may predict incorrect properties of  hadronic final states. It is therefore necessary to come up with independent validation methods, which would allow specific physical mechanisms to be tested.

The method developed in this paper is based on comparison with inclusive electron-scattering data.  Of course, there are a number of differences with neutrino scattering, the most important one being the absence of the axial contribution at the primary interaction vertex. Nevertheless, several crucial factors are common, such as the framework for various hadronic effects and the nuclear model.
What electron scattering offers is the advantage of precisely specified kinematics: the initial and final electron energies are accurately known and high-statistics samples can be accumulated at fixed scattering angles. In many situations, this allows one to identify problems in specific physical processes.

In our investigation, we consider \GENIE{}~\cite{Andreopoulos:2009rq,Andreopoulos:2015wxa,genieurl}, primarily because this generator is at present employed by all Fermilab-based neutrino experiments. The \GENIE{} mission statement explicitly stresses the importance of modeling electron-nucleus interactions in the same physics framework as neutrino-nucleus interactions. Insofar as this crucial principle is adhered to in the code, electron scattering data should provide an excellent validation framework.  It is also important that the generator has so far not been tuned to such data. We will return to the relationship between electrons and neutrinos in \GENIE{} repeatedly throughout this paper.\phantom{\cite{Katori:2013eoa}}

\begin{figure*}
\centering
    \subfigure{\label{fig:Barreau_a}}
    \subfigure{\label{fig:Barreau_b}}
    \subfigure{\label{fig:Barreau_c}}
    \includegraphics[width=0.90\textwidth]{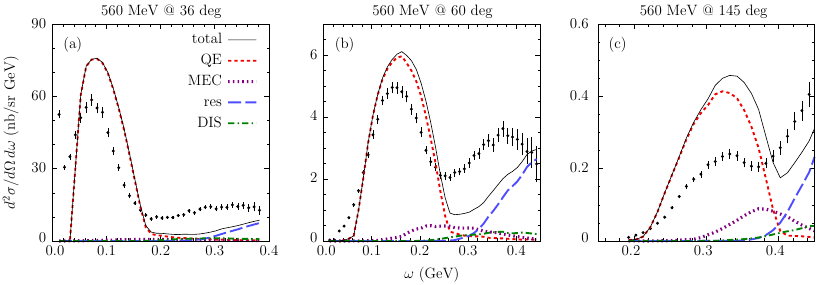}
\caption{\label{fig:Barreau} Comparisons of the \genie{} predictions for the double differential cross section for electron scattering off carbon with the data reported by Barreau \etal{}~\cite{Barreau:1983ht}.}
\end{figure*}

To set up the problem, let us consider an earlier test of \GENIE{}~\cite{Katori:2013eoa} against electron-carbon scattering data from Ref.~\cite{Barreau:1983ht}. That test received much attention at the time, as it revealed large disagreement between the generator predictions and the data. Since the comparison employed an old version of \GENIE{}, one might question if its findings are still relevant. In fact, we find very similar results with a recent \GENIE{} version\footnote{We thank Steven Dytman for detailed help on how to run \GENIE{} in the electron scattering mode.}. This is demonstrated in Fig.~\ref{fig:Barreau}, where we repeat the comparisons~\cite{Katori:2013eoa}, but with version 2.12, which, as of this writing, is still being used by most neutrino experiments. The generator predictions are seen to be very similar to those in Ref.~\cite{Katori:2013eoa}, up to sampling issues in the original paper. Thus, the problem remains as timely as ever.

The test~\cite{Katori:2013eoa} was limited to an initial electron energy of 560 MeV and large scattering angles, from $36^\circ$ to $145^\circ$. One might wonder if the issues identified by it were specific to those kinematic conditions, optimized for searching for multinucleon effects. Do the discrepancies extend to beam energies above 1~GeV, relevant for DUNE? And if there are still discrepancies at DUNE energies, can one identify the physical processes behind them?

The goal of the present paper is to carry out a systematic electron-scattering comparison of \GENIE{} in the kinematic regimes relevant to DUNE and NOvA. This will allow us to go beyond identifying discrepancies with individual datasets, and map out the patterns of discrepancies across datasets. Such patterns can then be used to guide generator improvements.

The broad coverage of the space of kinematic conditions will allow us to address our second main aim: to establish in which physical regimes the generator discrepancies are most severe. Experiments such as NOvA and MINERvA have been focusing much of their recent cross section studies on multinucleon effects, specifically on the so-called meson-exchange currents (MEC). Given the overall richness of physics, it is not {\it a priori} obvious that all the other processes are successfully dealt with in the generators. Indeed, as we will see, they are not.

Once the patterns of discrepancies are identified, our next task will be to establish their physical origin. We will do this by examining the same kinematic regimes in other---simpler---nuclei. This will make it possible to decisively disentangle fundamental neutrino-nucleon interaction effects from those created by the presence of the nuclear medium.

Our presentation is organized as follows. In Sec.~\ref{sect:DUNEregime}, we define the range of energy- and momentum-transfer values relevant for DUNE and NOvA. In Sec.~\ref{sect:kinematicwindow}, we show how in this regime precise kinematics of electron-nucleon scattering allows one to clearly separate specific physical processes in the data. In Sec.~\ref{sect:gutsofgenie}, we briefly summarize the models implemented in \genie{} to treat these physical processes. With the stage set, we confront \genie{} predictions with electron scattering data. In Sec.~\ref{sect:firstlook}, we consider recent datasets collected for a few nuclear targets at the same kinematics; we find that \genie{} reproduces certain features of the data quite well, while dramatically mispredicting certain other features.  In Sec.~\ref{sect:Carbonmapping} we systematically investigate these discrepancies across the DUNE and NOvA kinematic regime with world's inclusive carbon data. This is followed by the comparisons in Sec.~\ref{sect:diagnosis} to the deuteron and hydrogen data, which allows for the decisive diagnosis of the origin of the main discrepancies. In Sec.~\ref{sect:discussion}, we summarize and organize our main findings and discuss the synergies between neutrino and electron scattering experiments. Our final summary and overall thoughts are presented in Sec.~\ref{sect:conclusions}.

The details of the data used in this comparison are supplied in Appendix~\ref{appendix:data}. In Appendix~\ref{appendix:DISReduction}, we provide an example of empirical tuning that can be immediately implemented based on the results of our electron scattering comparison and discuss its advantages and limitations. In Appendix~\ref{appendix:CLASdata} we acknowledge experiments that did not report their measurements in form of cross sections, urging them to follow through with this important step of data analysis.\newpage

\section{Scattering regimes of DUNE~and~\nova{}}
\label{sect:DUNEregime}
As already mentioned, this paper extends electron-scattering tests to the kinematic regimes of DUNE and \nova{}. Let us begin by defining these regimes.

\begin{figure}
\centering
    \includegraphics[width=0.85\columnwidth]{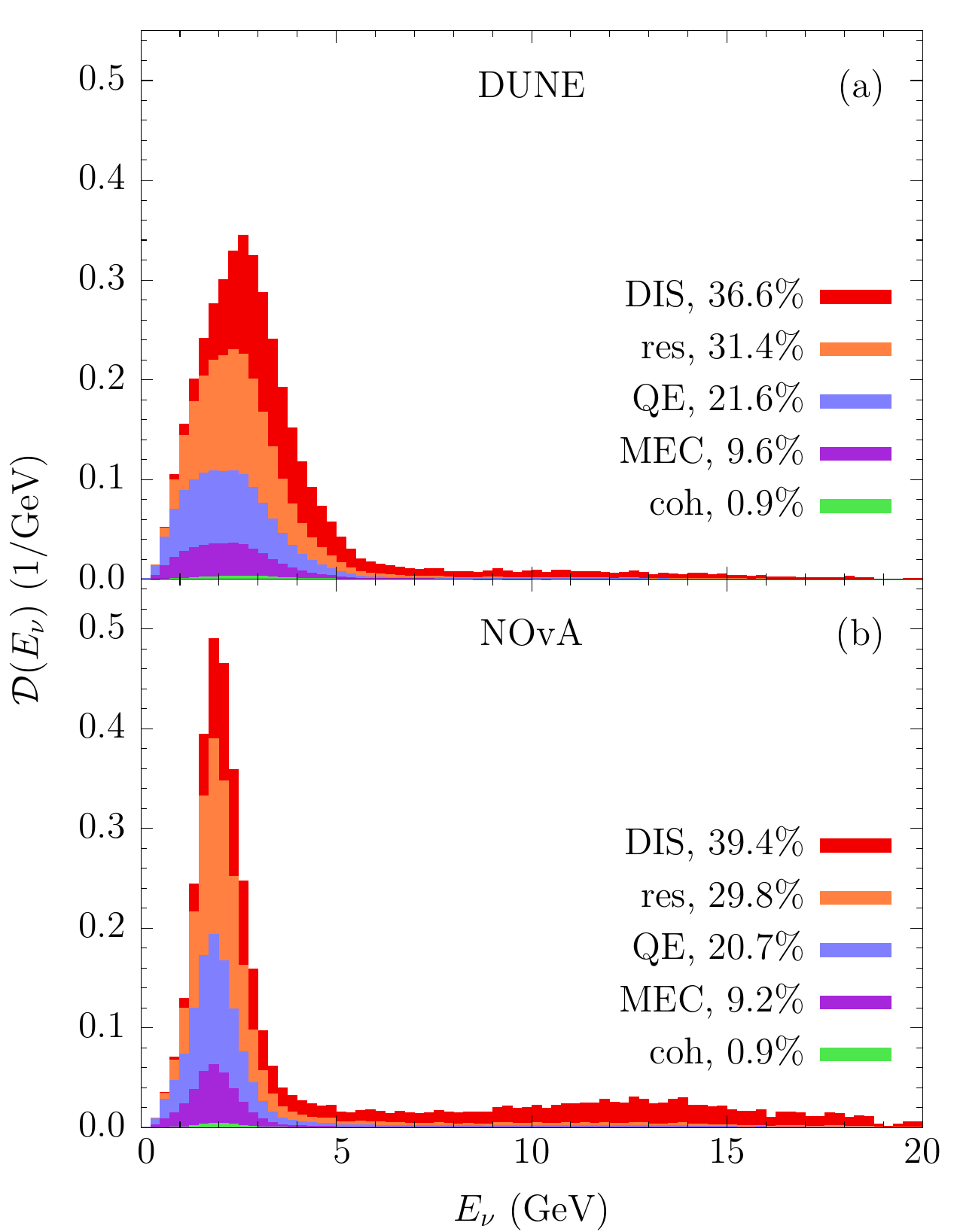}
\caption{\label{fig:spectra} Charged-current event distributions as a function of neutrino energy obtained using \genie{} for muon neutrino scattering off carbon for the beam fluxes at the near-detector sites of (a) DUNE~\cite{DUNE_flux} and (b) \nova{}~\cite{NOvA_flux}, shown as stacked histograms.}
\end{figure}

In order to achieve sensitivity to mass hierarchy, oscillation experiments can take advantage of matter effects in the Earth, which become important on distance scales $\sim$1,000 km (see, e.g., Ref.~\cite{Friedland:2018vry} for further discussion). Given the measured value of the atmospheric mass-squared splitting parameter, $\Delta m_{23}^2 \simeq 2.5 \times 10^{-3}$ eV$^2$, a baseline of the order of 1,000 km corresponds to the strongest $\nu_e$ appearance signal at energies $\sim$3 GeV. In the case of DUNE and \nova{}, due to different baselines---1,300 and 810 km, respectively---the beam fluxes are peaked at somewhat different energies in the 2--3 GeV region. We depict the event spectra calculated for both beams in Fig.~\ref{fig:spectra}.

\begin{figure}[hb]
\centering
    \includegraphics[width=0.85\columnwidth]{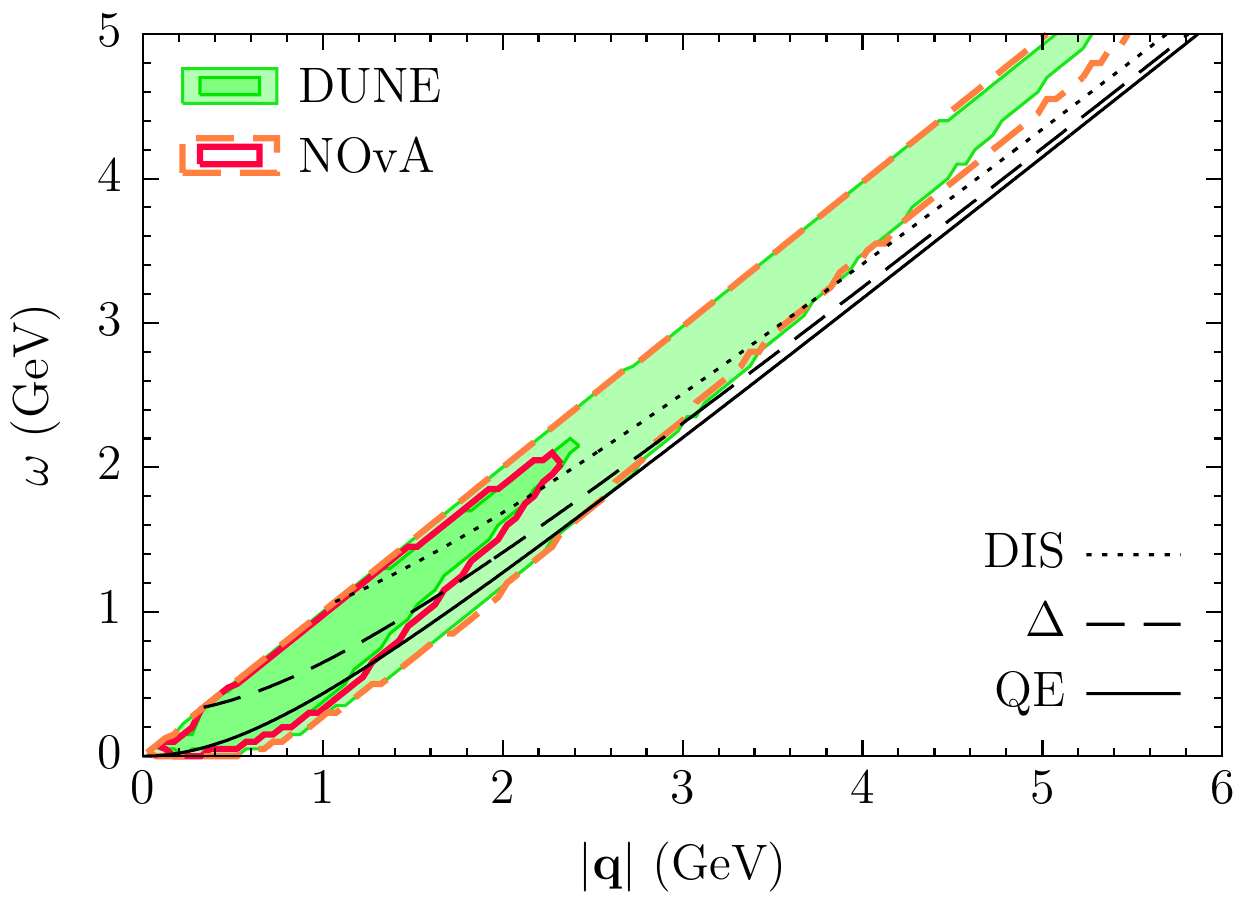}
\caption{\label{fig:DUNE_vs_NOvA} Comparison of the regions corresponding to 68 and 95\% of charged-current events in DUNE (dark and light shaded areas) and in NOvA (thick solid and dashed lines) according to \genie{}. The regions largely overlap between the two experiments. The thin solid, dashed, and dotted lines present the kinematics corresponding to quasielastic scattering, $\Delta$ excitation, and deep inelastic scattering at $W=1.7$ GeV on free nucleons.}
\end{figure}

As evident from this figure, the two beams have different widths. The \nova{} beam is designed to be narrow-band, which is achieved by placing the detectors off axis. This choice is made to focus on the first oscillation maximum, while minimizing feed-down of neutral-current events from higher energies. In contrast, DUNE will use a broad-band beam, with the aim of mapping out the oscillation probability over a range of energies, down to the second oscillation maximum. This is achieved by placing the detector on axis.
At the same time, however, the \nova{} beam does have a more-pronounced high-energy tail than DUNE. This is a consequence of the third horn in DUNE beamline, which is absent in the \nova{}'s configuration.

The net result is that, for the purpose of our cross section comparison, the kinematics of DUNE and \nova{} turn out to be quite similar. The desired kinematic window is depicted in  Fig.~\ref{fig:DUNE_vs_NOvA}, in the plane of momentum transfer $\n{q}\equiv|\mathbf{k}_i-\mathbf{k}_f|$ and energy transfer $\omega=E_i-E_f$, the quantities that can be directly obtained from the initial $i$ and final $f$ lepton's momenta and energies.
The regions corresponding to 68\% and 95\% of muon-neutrino events produced in charged-current scattering off carbon are presented as the filled regions for the DUNE beam and as the closed contours for the \nova{} beam. The regions are obtained assuming perfect detection of muons, to keep our considerations independent of specific estimates of detector effects.

Henceforth, whenever we refer to the kinematics of DUNE, it should be understood that our findings apply, to almost the same extent, to the kinematics of \nova{}. In fact, because of the scarcity of  electron-scattering data for argon, we focus our most extensive analysis on the carbon target. This should make our findings directly relevant for the estimates of systematic uncertainties in \nova{}, the detectors of which are composed predominantly of carbon.\phantom{\cite{Bloom:1971ye}}

\begin{figure*}
\centering
    \subfigure{\label{fig:C_data_a}}
    \subfigure{\label{fig:C_data_b}}
    \includegraphics[width=0.90\textwidth]{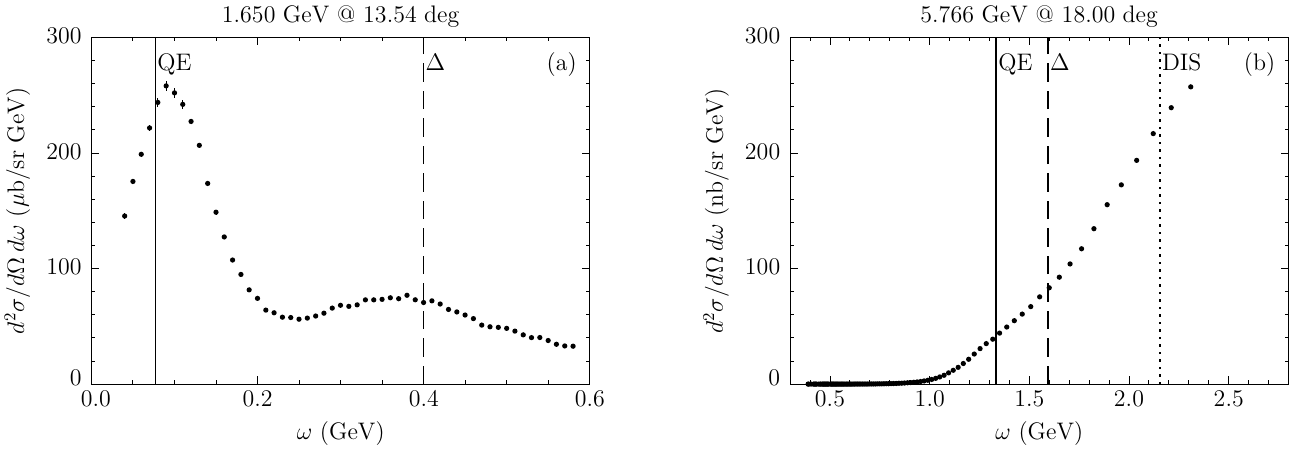}
\caption{\label{fig:C_data} Double differential cross section for electron scattering off carbon according to the measurements reported in (a) Ref.~\cite{Baran:1988tw} and (b) Refs.~\cite{Fomin:2008iq,Fomin:2010ei}. The horizontal solid, dashed, and dotted lines correspond to the kinematics of quasielastic interaction, $\Delta$ excitation, and the onset of deep-inelastic scattering on free nucleons, respectively.}
\end{figure*}

Figure~\ref{fig:DUNE_vs_NOvA} also shows curves corresponding to different scattering processes, as labeled in the legend on the right. We will describe this physics next.\newpage

\section{Scattering physics in the DUNE kinematic window}
\label{sect:kinematicwindow}

As we have seen, neutrino interactions at DUNE are characterized by momentum- and energy-transfer values ranging from sub-GeV to $\sim$5 GeV and even beyond. In this kinematic window, one has to model both QE scattering, in which a struck nucleon remains unbroken, and pion-producing processes. Pion production can occur through the excitation of baryonic resonances, such as $\Delta(1232)$ and higher states, or through nonresonant channels. At high energy and momentum transfers, the DIS picture becomes appropriate, with the primary vertex treated at the quark level, followed by a hadronization process. The locations of these main processes is indicated in Fig.~\ref{fig:DUNE_vs_NOvA} by the thin solid, dashed, and dotted curves, as marked in the legend.

The curves are meant to guide the eye, with the actual physical processes distributed in continuous bands around them. Several factors contribute to this broadening of the features. For example,
the curves are drawn for nucleons at rest, while, in a nucleus---such as carbon, argon, or iron---nucleons experience Fermi motion. Also potentially important are more subtle effects of nucleon binding energy, nucleon pairing, and multinucleon currents that we will return to in some detail later.

Following the convention in \genie{}, we denote the value of the parameter $W=1.7$ GeV as the onset of the DIS regime. This parameter is defined as the invariant mass of the final-state hadronic system, $W^2=M^2 + 2 M \omega - Q^2$, where $M$ is the nucleon mass, $\omega$ is the energy transfer, as before, and $Q^2\equiv\n{q}^2-\omega^2$. In reality, the transition from the resonant regime to DIS is gradual and, indeed, in the intermediate region, a dual description in terms of quarks or hadrons should be possible~\cite{Bloom:1971ye}.

In electron-scattering data, in many cases these features can be seen directly, as illustrated in Fig.~\ref{fig:C_data}. In the left panel, we immediately recognize the quasielastic peak, broadened by Fermi motion and shifted by the effect of nucleon binding. At higher energy transfers, we also see a broad feature created mostly by the $\Delta(1232)$ resonance. The high-$\omega$ tail of the dataset gradually transitions to the DIS regime (the $W=1.7$ GeV marker would lie outside of the plotted range of $\omega$.) This dataset was collected for the electron beam of 1.65 GeV, at the scattering angle of 13.54\degree~\cite{Baran:1988tw}.

In the right panel, we show the data collected for a higher energy, 5.766 GeV, and scattering angle 18.00\degree~\cite{Fomin:2008iq,Fomin:2010ei}. In this case, a qualitatively different scenario is realized: inelastic hadronic processes become dominant, with the presence of the quasielastic bump barely detectable. The transition between the ``$\Delta$-resonance" and ``DIS" markers is, once again, continuous and featureless.

\begin{figure}[ht]
\centering
    \subfigure{\label{fig:events_2D_a}}
    \subfigure{\label{fig:events_2D_b}}
    \includegraphics[width=0.85\columnwidth]{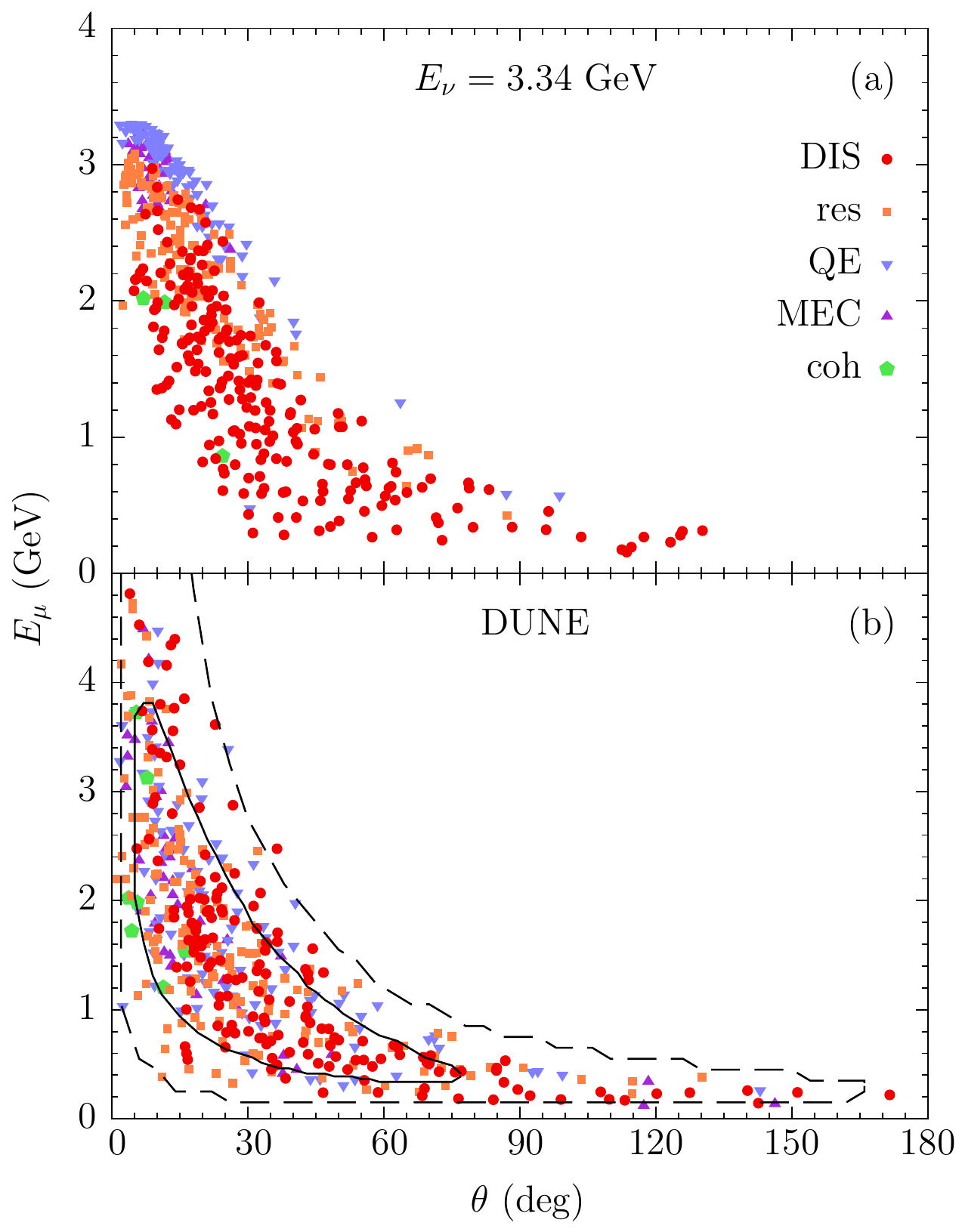}
\caption{\label{fig:events_2D} Scatterplot of the charged-current $\nu_\mu$ event distribution as a function of muon's production angle and its energy, obtained using \genie{} for (a) the average beam energy and (b) the near-detector flux of DUNE~\cite{DUNE_flux} and the carbon target. In panel (b), the solid (dashed) line shows the region corresponding to 68\% (95\%) of the events. }
\end{figure}

Can different scattering regimes be likewise delineated in the case of neutrino-nucleus scattering? To answer this question, we turn to Fig.~\ref{fig:events_2D}, where we present scatterplots of the final muon kinematics, created in the process $\nu_\mu+ \text{C} \rightarrow \mu + X$. The vertical axis shows the muon energy, while the horizontal axis presents its angle with respect to the neutrino beam. In the top panel, we plot the hypothetical case of a monochromatic neutrino beam, with $E_\nu=3.34$ GeV, the expected average energy of (unoscillated) events in the DUNE beam. This case shares many features with the electron scattering examples of Fig.~\ref{fig:C_data}: the transition between the resonant and DIS events is also gradual, and the quasielastic events are concentrated in a strip. In the actual DUNE beam, the situation is quite different, however, as depicted in the bottom panel. We can see that after taking into account the broad spectrum of neutrinos, the different interaction channels cannot be clearly separated by using the muon kinematics alone.

This example provides an effective illustration of the advantage one gains by using electron-scattering data. Precise control over the scattering kinematics it affords makes it possible to separate the different scattering regimes. This separation may provide a way to pinpoint specific physical processes behind any discrepancies. We will take advantage of this later in the analysis.

\section{Monte Carlo generator \genie{}}
\label{sect:gutsofgenie}

Figure~\ref{fig:events_2D} is an example of the simulation output of \GENIE{}, in which events are labeled by physical processes the generator invokes. In addition to the deep-inelastic scattering (``DIS''), resonance (``res'') excitation, and quasielastic (``QE'') interactions, we also see two more categories, meson-exchange currents (``MEC'') and coherent (``coh''). Let us now briefly summarize the physics models the code employs for each process.

From the outset, we note that for our studies here we use \genie{} version 2.12.10, with all settings set to their default values. This choice is deliberate: (i) this version is most frequently used in experimental simulations and (ii) we avoid tuning or otherwise altering the run settings to make the test as unbiased as possible.

For quasielastic scattering, the nucleon form factors are taken from the global fits~\cite{Bradford:2006yz,Kuzmin:2006dh}, and
the default nuclear model is the relativistic Fermi gas (RFG) model of Bodek and Ritchie~\cite{Bodek:1980ar}. This model takes into account the Fermi motion of the nucleons, while also adding a high-momentum tail---inspired by the effects of short-range correlations between nucleons---above the Fermi momentum, $p_F\simeq220$ MeV. The separation energy is assumed to be momentum independent and fixed to a value $\sim$25 MeV, chosen to reproduce the position of the QE peak in electron scattering. While in the model of Bodek and Ritchie interacting nucleons form deeply bound states, this feature is not implemented in \genie{}, in which nucleons in the high-momentum tail are typically unbound.

This treatment implicitly relies on the assumption that the process of scattering can be described as involving predominantly a single nucleon in the nucleus, with the remaining nucleons acting as a spectator system. This scheme is valid when the spatial resolution of the probe---of the order of $1/\n q$---is higher than the typical distance between nucleons in a nucleus, $\sim$1/(0.2 GeV). Under this condition, interference of the scattering amplitudes from different nucleons can be neglected. Likewise, one assumes that the duration of the interaction---of the order of $1/\omega$---is shorter than the timescale on which nucleons can appreciably interact with the rest of the nucleus, the so-called plane-wave impulse approximation.

The single-nucleon framework underlies the treatment of not only quasielastic scattering, but also different mechanisms of pion production, as described below. One process, for which it is manifestly violated, is coherent scattering. This process occurs when momentum transfer is small enough that interaction occurs on the nucleus as a whole. Another is the phenomenon of collective nuclear excitations, relevant at $\n q\lesssim100$ MeV, involving long-range correlations between nucleons.

Still another phenomenon involving more than one nucleon is the process of the meson-exchange currents mentioned before. In the default configuration of \genie{} version 2.12, it is described using the empirical procedure of Dytman~\cite{Katori:2013eoa}, which itself is a modification and extension to neutrino interaction of the prescription of Ref.~\cite{Lightbody:1988gcu}, developed for electron scattering. The invariant mass of two-nucleon events is assumed to have a Gaussian distribution centered at the value $W=(M+M_\Delta)/2$, the average of the nucleon and $\Delta$-resonance masses. The lepton kinematics is distributed according to the magnetic contribution to the elementary cross section. The strength is set to be a constant fraction of the elementary cross section for QE scattering, and to exhibit linear dependence on the mass number $A$.

The treatment of pion production in \genie{} depends on the value of the invariant mass of the final-state hadronic system, $W$. For $W\leq1.7$ GeV, \genie{} considers excitation of nucleon resonances using the framework of the Rein--Sehgal model~\cite{Rein:1980wg}. Compared with 18 in the original work, 16 resonances of unambiguous existence are implemented using up-to-date parameters, but neglecting any interference between them. Specifically, the form factors for the relevant transitions, including the axial parts, are taken from the global fits in Refs.~\cite{Kuzmin:2006dh}. In charged-current neutrino interactions, the effect of the charged lepton's mass is taken into account only in the calculations of phase-space boundaries.

Meson-production processes not involving resonance excitation, referred to in \genie{} as deep-inelastic scattering, are modeled following the method of Bodek and Yang of Refs.~\cite{Bodek:2002ps,Bodek:2004pc}. This effective approach relies on leading-order parton-distribution functions~\cite{Gluck:1998xa}, and introduces effective masses of the target and final state to account for higher-order corrections and to extend applicability of the parton model to the low-$Q^2$ region. For $W\geq1.7$ GeV, DIS is the only mechanism of interaction included in \genie{}. In the resonance region, $W<1.7$ GeV, DIS is employed to produce a nonresonant background of events involving one or two pions. It should be stressed that the Bodek-Yang model is conceived by its authors to capture all inelastic physics beyond the $\Delta$ resonance. We will revisit this issue in Sec.~\ref{sect:diagnosis}.

Although in this analysis we do not tackle the problem of hadronization, it is worth noting that in \genie{} it is performed relying on the approach of Ref.~\cite{Yang:2009zx}. At the invariant hadronic masses below 2.3 GeV, a phenomenological prescription, based on Koba--Nielsen--Olesen (KNO) scaling~\cite{Koba:1972ng}, is employed. At $W>3$ GeV, hadronization is modeled with \pythia{} 6~\cite{Sjostrand:2006za}. Over the intermediate region $2.3\leq W\leq 3$ GeV, \genie{} linearly transitions between the two hadronization models.

As a final remark of this section, we want to reiterate that the design of \GENIE{} makes it possible to treat electron-nucleus and neutrino-nucleus interactions using use the same physics framework~\cite{Andreopoulos:2009rq,Andreopoulos:2015wxa,genieurl}. This feature has been made use of in the past to verify the overall consistency
of the \genie{} physics model and the DIS implementation by comparisons to electron scattering data, as reported in Ref.~\cite{Andreopoulos:2009rq}. Although the code of version 2.12 does not connect the description of electron and neutrino interactions as fully as the one of version 3~\cite{Roda:2017ovf}, even in version 2.12 the instances where these processes are implemented separately are rather an exception than a rule. We will return to this issue in Sec.~\ref{sect:discussion}.

\section{\GENIE{} predictions vs. GeV electron scattering data: first look}
\label{sect:firstlook}

\begin{figure}[t]
\centering
    \includegraphics[width=0.85\columnwidth]{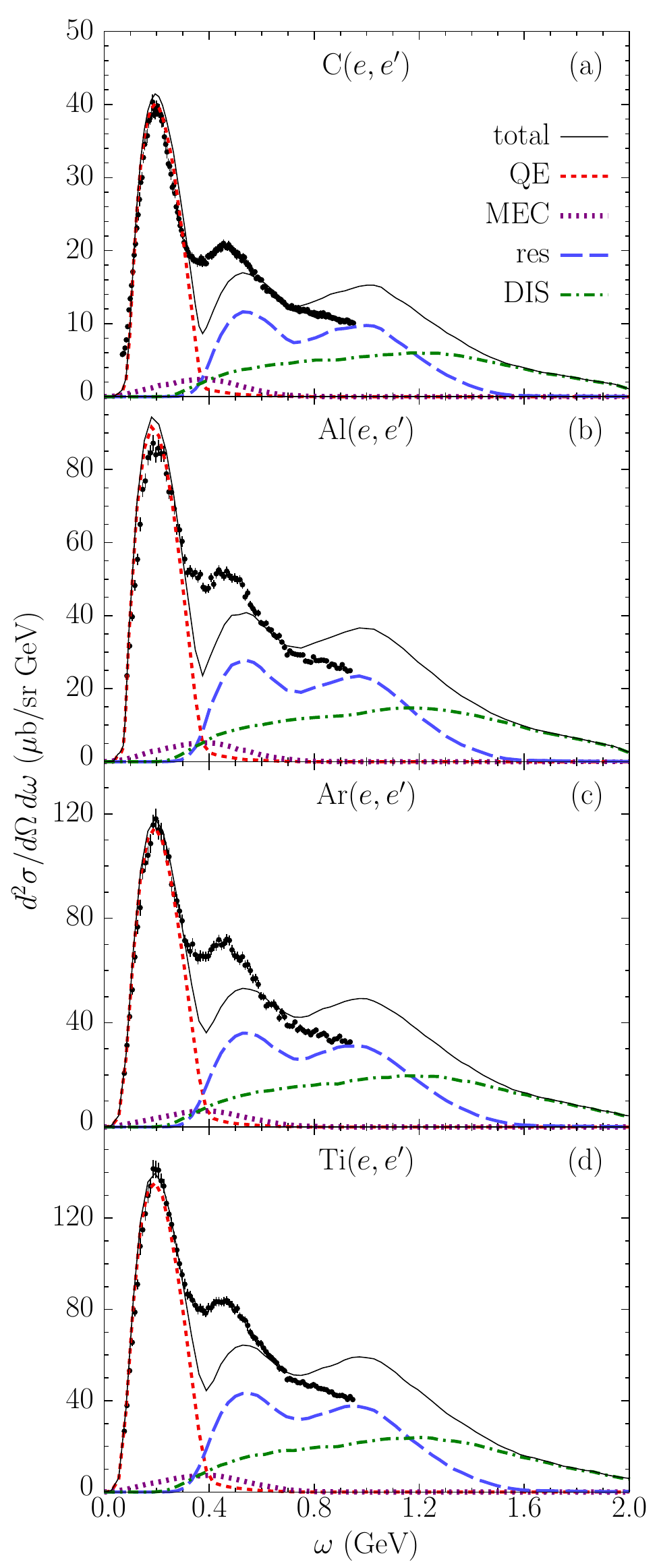}
\caption{Double differential cross sections for electron scattering off (a) carbon, (b) aluminum alloy 7075, (c) argon, and (d) titanium at beam energy 2.222-GeV and scattering angle $15.54^\circ$. Predictions of \genie{} are compared with recent JLab measurements reported in Refs.~\cite{Dai:2018xhi,Dai:2018gch,Murphy:2019wed}.}
\label{fig:CTi}
\end{figure}

We have already seen in Fig.~\ref{fig:Barreau} that for sub-GeV energies these models, as implemented in \genie{}, show significant discrepancies with electron scattering data. Let us now consider electron beams in the few-GeV energy range. As our first illustration, we will use inclusive scattering datasets recently collected at Jefferson Laboratory (JLab) using a 2.2 GeV electron beam~\cite{Dai:2018xhi,Dai:2018gch,Murphy:2019wed}. An important advantage of these measurements is that they include argon and titanium---making them directly relevant to DUNE---as well as carbon and a target made of aluminum alloy 7075, allowing for simultaneous investigations across a range of nuclei.

Figure~\ref{fig:CTi} shows comparisons of the data taken at the scattering angle of  $15.54^\circ$ to the corresponding predictions of \genie{}. The results are presented as a function of energy transfer $\omega=E_i - E_f$, where, as before, $E_i$ is the beam energy and $E_f$ is the final electron energy. The corresponding values of momentum transfer $\n q =|\ve k_e - \ve k_e'|$, with $\ve k_e$ and $\ve k_e'$ being the electron's initial and final momenta, increase monotonically from 0.60 to 1.05 GeV for $\omega$ between 0.07 and 0.95 GeV. For such momentum transfers, the process of scattering off a nucleus can be described within the framework discussed in the previous section, with most interactions involving a single nucleon in the nucleus.

Just like in our discussion in Sec.~\ref{sect:kinematicwindow}, in Fig.~\ref{fig:CTi} one can clearly distinguish the different scattering mechanisms at work: the QE peak at $\omega\sim0.2$ GeV, the $\Delta$-excitation peak at $\omega\sim0.45$ GeV, and the high-$\omega$ tail of the cross section receiving contributions from the excitation of higher resonances. The features of these structures---width, position, shape, and height---provide information on the distribution of the momentum and energy of the struck nucleons, including their binding energy in the nucleus, as well as on the properties of the elementary electron-nucleon vertex. The resulting cross sections are combinations of these factors and it is not immediately possible to unfold them. This task requires analyzing much more data and will be the goal of the rest of this paper.

It is, however, already possible to make a number of relevant observations, based on the salient features of the four plots.
\begin{itemize}
    \item[(i)]  As can be seen from comparing Fig.~\ref{fig:CTi} with Fig.~\ref{fig:Ar_nu&anu} later in this paper, at these energies, the cross sections for electron scattering are some eight orders of magnitude greater than those for neutrino interactions. As a consequence, electron-nucleus cross sections can be measured with very small uncertainties in a short time. For example, the data in Fig.~\ref{fig:CTi} were collected over less than an hour for each of the targets.
    \item[(ii)] Using electrons enables one to collect scattering data over a broad kinematic region of relevance for neutrino experiments. The virtual photon mediating the interaction can carry a variety of $\n q$ and $\omega$ values. This contrasts with photonuclear scattering, in which the real photon is limited to $Q^2=0$.
    \item[(iii)] The \genie{} generator does a remarkably good job describing the quasielastic scattering regime at this kinematics. Both the location of the QE peak and the width of the feature are in good agreement with the data.
    \item[(iv)] The agreement in the QE regime is good both for the argon and titanium targets. This has important implications for the expected generator performance in the (anti)neutrino mode. Indeed, while QE antineutrino scattering involves initial-state protons, QE neutrino scattering involves initial state neutrons. Thus, we need information on the distribution of both in the argon nucleus. Fortunately, the information on neutrons can be deduced from electron scattering off titanium $_{22}$Ti, the proton structure of which \emph{mirrors} the neutron one of $^{40}_{18}$Ar. The results for titanium and argon are presented in the bottom two panels. The predicted position of the QE peak agrees with the data up to $\sim$10 MeV for titanium and $\sim$5 MeV for argon.
    \item[(v)] The situation, however, is dramatically different to the right of the QE peak. The cross sections calculated using \genie{} overestimate the data by up to $\sim$40\%  in the region of high energy transfers, where both DIS and excitation of higher resonances contribute, and underestimate the $\Delta$ production peak by $\sim$30\%. These discrepancies may at first escape detection in comparisons to neutrino data, in which they could be washed out upon integration over the beam spectrum, but they are glaring in electron scattering.
    \item[(vi)] In the so-called dip region---between the QE and $\Delta$ peaks---the electron data are underestimated by up to $\sim$40\%. One's first response to this deficit might be to increase the MEC component. Notice, however, that the MEC contribution to the \genie{} cross section extends under the QE peak, which broadens it, increases its height, and makes it more asymmetric. Simply increasing the normalization of the MEC component would spoil the remarkable agreement with the data in the QE region we discussed above.
    \item[(vii)] The pattern of deficits and excesses is consistent across the four nuclear targets. In all cases, the QE regime is reproduced well, the $\Delta$ peak is shifted to high energy transfers, the dip region rate is underpredicted, and the rate in the DIS transition region is dramatically overpredicted.
\end{itemize}

The last point proves to be very important for the next step of our investigation. Having found large discrepancies in the regime of inelastic hadronic interactions, our task is to systematically map out these discrepancies across the relevant kinematic space and, eventually, to understand their physical origin. Ideally, one would wish to map out the kinematic space with argon and titanium measurements. Unfortunately, such data are currently not available. In addition to Refs.~\cite{Dai:2018xhi,Dai:2018gch}, the only other measurement at the kinematics of relevance to DUNE was reported by Anghinolfi {\it et al.}~\cite{Anghinolfi:1995bz}. However, the scaling analysis performed in Ref.~\cite{Dai:2018gch} revealed issues with the data~\cite{Anghinolfi:1995bz} and, therefore, we do not discuss them here. We are, however, able to exploit the fact that exactly the same discrepancies are seen with the carbon dataset. This is extremely fortunate, for there is a large body of available electron-carbon scattering data, allowing us to carry out a systematic study and to draw robust conclusions. We present this study in the next section.
\phantom{\cite{Benhar:2006er}}

\section{Systematic tests of the \genie{}'s $(e,e')$ cross sections against\\ world's carbon data}
\label{sect:Carbonmapping}

Our next goal is to investigate whether the large discrepancies observed in Fig.~\ref{fig:CTi} for specific kinematics are part of a bigger pattern. Establishing such patterns in the space of energy and momentum transfers would be an important step toward identifying the sources of the discrepancies. For this investigation, we will use the carbon nucleus, which is by far the most extensively studied target in electron-scattering measurements~\cite{Benhar:2006er}.
Using available literature, we compiled measurements, accumulated over the last five decades, that span a broad range of beam energies and scattering angles~\cite{Whitney:1974hr,Barreau:1983ht,O'Connell:1987ag,Bagdasaryan:1988hp,Baran:1988tw,Sealock:1989nx,Day:1993md,Gomez:1993ri,Arrington:1995hs,Arrington:1998hz,Arrington:1998ps,Fomin:2008iq,Fomin:2010ei,Dai:2018xhi}. The details of the datasets used are summarized in Table~\ref{tab:summary_C} and discussed in Appendix~\ref{appendix:data}.

\begin{figure}
\centering
    \subfigure{\label{fig:comparisons_C_a}}
    \subfigure{\label{fig:comparisons_C_b}}
    \subfigure{\label{fig:comparisons_C_c}}
    \includegraphics[width=0.85\columnwidth]{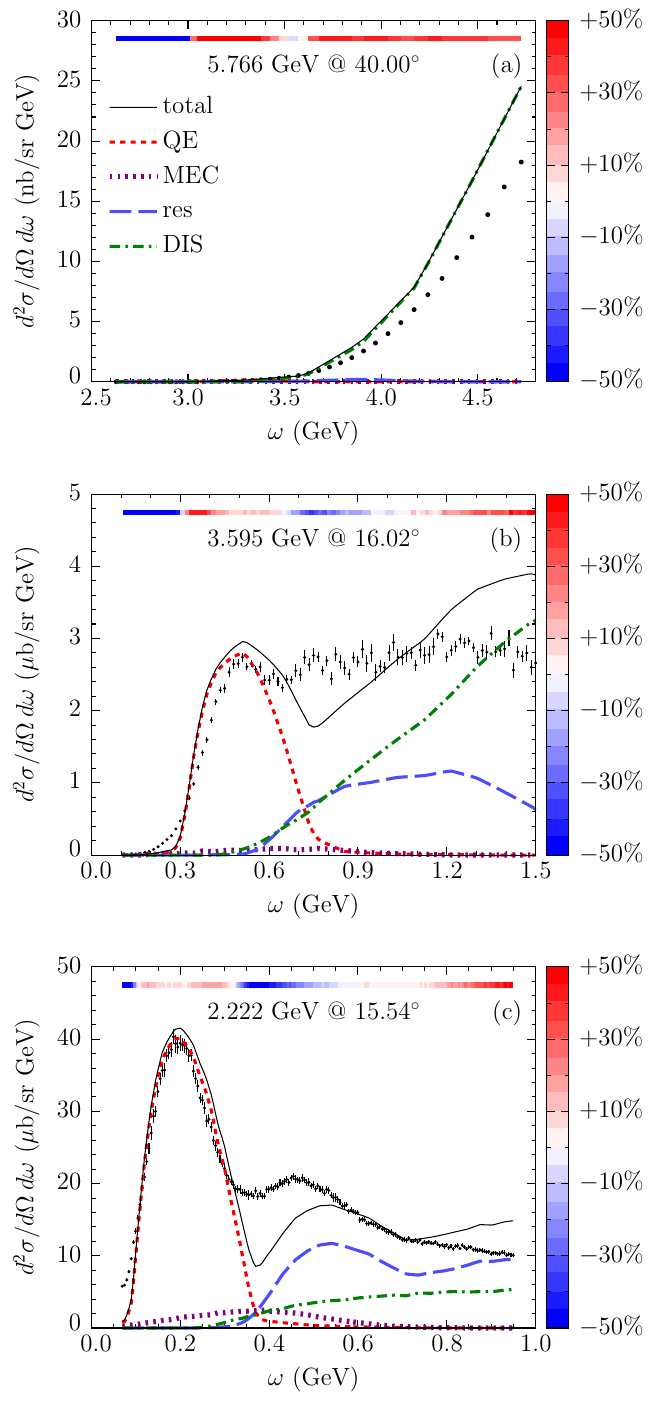}
\caption{\label{fig:comparisons_C} Comparisons of the predictions of \genie{} for the double differential cross section for electron scattering off carbon with the data~\cite{Fomin:2008iq,Fomin:2010ei,Day:1993md,Dai:2018xhi}. The values of the fractional difference between the calculations and data are represented as the one-dimensional heat maps on the top of each panel.}
\end{figure}

We start our presentation with a comparative analysis of the measurements reported in Refs.~\cite{Fomin:2008iq,Fomin:2010ei,Day:1993md,Dai:2018xhi}. They are shown in Fig.~\ref{fig:comparisons_C}, along with the corresponding \genie{} predictions, both the total results (solid curves) and the contributions of individual physical processes (see legend). This set of figures allows us to investigate how the findings of the last section (reproduced in the bottom panel for convenience) extend to higher energies and larger scattering angles (top two panels). This allows us to cover a range of scattering regimes relevant to DUNE.

\begin{figure*}
\centering
    \subfigure{\label{fig:DUNE_wideRange_a}}
    \subfigure{\label{fig:DUNE_wideRange_b}}
    \includegraphics[width=0.90\textwidth]{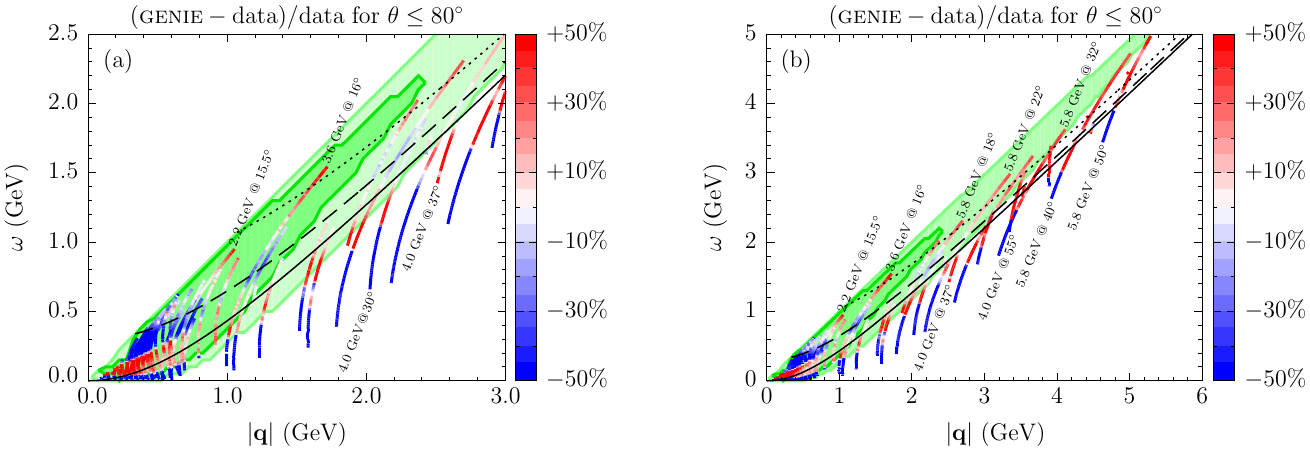}
\caption{\label{fig:DUNE_wideRange} Fractional difference between the \genie{} calculations and the experimental electron-carbon cross sections for momentum transfers extending to (a) 3 GeV and (b) 6 GeV, presented on the $(\n q, \omega)$ plane. The region corresponding to 68\% (95\%) of charged-current events in DUNE is shown as the dark (light) shaded area as in Fig.~\ref{fig:DUNE_vs_NOvA}.}
\end{figure*}

Let us first examine the QE peak, which is clearly visible in Figs.~\ref{fig:comparisons_C_b} and \ref{fig:comparisons_C_c}. Overall, it is perhaps the best reproduced feature in both cases, although on closer examination of Fig.~\ref{fig:comparisons_C_b} there are some interesting and subtle details. In the range of $\omega$ between 0.3 and 0.5 GeV, the RFG model overpredicts the cross section, while below 0.3 GeV it significantly underpredicts it. Several nuclear effects can \textit{a priori} contribute to this behavior, including (i) scattering on nucleon pairs~\cite{Heidmann:1950zz,Levinger:1951vp}; (ii) depletion of single-particle nuclear levels due to the existence of nucleon pairing; (iii) shift of the QE peak due the modified  nucleon dispersion relation in nuclear matter~\cite{Benhar:2006wy}.
The combination of these effects creates deviations from the RFG peak profile, in a way that can be consistently modeled~\cite{Ankowski:2014yfa}. Yet, overall, their impact is subtle, compared to the size of discrepancies to the right of the QE peak.

One might be tempted to ascribe the entire deficit of the cross section seen in Fig.~\ref{fig:comparisons_C_b} at energy transfers $\sim$0.6--0.9 GeV, in the dip region, to the understated MEC component. This explanation, however, runs into difficulty with the shape of the QE peak. Indeed, if one were to fill the dip purely by tuning (increasing) the normalization of the MEC component, one would get large disagreement immediately to the right of the QE peak. This behavior was already noted in Sec.~\ref{sect:firstlook}, but at the higher energies of Fig.~\ref{fig:comparisons_C_b} it is even more pronounced. Even with the default MEC normalization used in \genie{}, the excess of the calculated peak height is in this case $\sim$10\%.

Although the cross section in the dip region is underestimated and the contribution of higher resonances is overestimated, at certain energy transfers above the $\Delta$-excitation peak in Fig.~\ref{fig:comparisons_C_c}, the prediction of genie agrees very well with the data. In particular, for $0.54\leq\omega\leq0.76$
GeV, the calculations do not differ from the experimental
results by more than 7\%. Later on, we will revisit this
issue, in view of results for other targets.

As one considers still larger $\omega$ values, the dominant contribution to the cross section comes from hadronic inelasticities, namely DIS or higher resonances. We see that the predicted cross sections in this regime show large discrepancies with the data in all three panels. Specifically, at the largest values of $\omega$, the calculated results overestimate the data by $\sim$45\% in Fig.~\ref{fig:comparisons_C_a}, $\sim$53\% in Fig.~\ref{fig:comparisons_C_b}, and $\sim$47\% in Fig.~\ref{fig:comparisons_C_c}. This suggests that the problematic generator behavior in the inelastic regime first seen in Sec.~\ref{sect:firstlook} is indeed a rule, rather than an exception.

To investigate this further, we will turn to the full set of data given in Table~\ref{tab:summary_C}, which makes it possible to cover the kinematic space of DUNE. To visualize our results, we will employ one-dimensional heat maps showing fractional differences between \genie{} calculations and electron-scattering data.
These heat maps are already used in Fig.~\ref{fig:comparisons_C}, where they appear as thin colored bands near the top of each panel. The color bars to the right of each panel show the legend, in terms of fractional differences from $-50\%$ to $+50\%$. Negative (positive) numbers imply that the calculation underestimates (overestimates) the data.
The differences exceeding $\pm50\%$ are included in the corresponding extreme colors.

For example, in Fig.~\ref{fig:comparisons_C_b}, for $\omega<0.3$ GeV, the deep blue region in the heat map indicates that the data are underestimated by more than 50\% (in fact, 100\%). This comes from the scattering off strongly interacting nucleon pairs, as noted above. In the region of $\omega$ between $0.3$ and $0.5$ GeV, up to the QE peak, the generator moderately overestimates the data, leading to a red segment in the color map. The red region continues beyond the peak position, to $\omega\sim0.6$ GeV, where the MEC contribution increases the difference between the calculations and data. Finally, for larger $\omega$ values, one has both blue and red segments, with absolute differences in the range of tens of percent. This is where the largest \emph{absolute} discrepancies occur.

\begin{figure*}
\centering
    \subfigure{\label{fig:DUNE_narrowRange_a}}
    \subfigure{\label{fig:DUNE_narrowRange_b}}
    \includegraphics[width=0.90\textwidth]{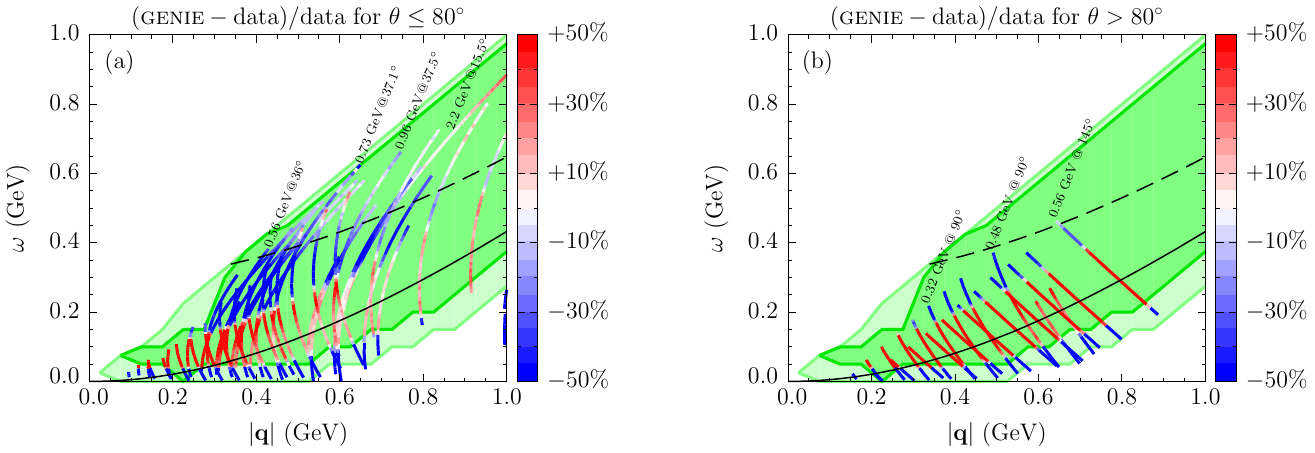}
\caption{\label{fig:DUNE_narrowRange} Same as Fig.~\ref{fig:DUNE_wideRange} but focusing on momentum transfers up to 1 GeV, and comparing the behavior of the data for scattering angles (a) up to 80 degrees and (b) exceeding 80 degrees. Note that no data for $\theta > 80^\circ$ are available for $\n q> 1$ GeV.}
\end{figure*}

The kinematic region of DUNE has been previously identified in Fig.~\ref{fig:DUNE_vs_NOvA}, in the plane of momentum transfer $\n q$ and energy transfer $\omega$. It is reproduced in Fig.~\ref{fig:DUNE_wideRange}, as the light and dark green shaded regions, containing correspondingly 68 and 95\% of charged-current events in DUNE. On top of these regions, we superimpose the heat maps of the comparisons of \genie{} predictions with the data listed in Table~\ref{tab:summary_C}.

As guidance on where different interaction mechanisms contribute in the $(\n q,\ \omega)$ plane, we use the $\n q$-dependence of the energy transfer in scattering on a free nucleon of mass $M$ at rest, leading to production of the hadronic state of the invariant mass $W$,
\[
\omega^\textrm{free}(W)=\sqrt{W^2+{\n q}^2} - M.
\]
In Fig.~\ref{fig:DUNE_wideRange} we show  $\omega^\textrm{free}(M)$, $\omega^\textrm{free}(\textrm{1.232 GeV})$, and $\omega^\textrm{free}(\textrm{1.7 GeV})$ as the solid, dashed, and dotted lines, respectively. These correspond to QE scattering, $\Delta$ excitation, and the onset of purely deep-inelastic scattering in \genie{}.

\begin{table}[hb]
\caption{\label{tab:summary_C} Summary of the cross sections extracted for inclusive electron scattering off carbon. Symbol ``Y'' marks the datasets considered in this analysis.}
\begin{ruledtabular}
\begin{tabular}{@{}lldddcc@{}}
    \multirow{2}{*}{Year} & \multirow{2}{*}{Lab} & \multicolumn{1}{c}{\textrm{Energy}} & \multicolumn{1}{c}{\textrm{Angle}} & \multicolumn{1}{c}{Point} & \multicolumn{1}{c}{Incl.} & \multirow{2}{*}{Ref.}\\
  &  & \multicolumn{1}{c}{\textrm{(GeV)}} & \multicolumn{1}{c}{\textrm{(deg)}} & \multicolumn{1}{c}{number} & \multicolumn{1}{c}{here} & \\
    \hline
    1974& HEPL  & 0.50 & 60.0 & 35 & Y &  \cite{Whitney:1974hr}\\
    1983& Saclay & \multicolumn{1}{c}{\textrm{0.12--0.68}} & \multicolumn{1}{c}{\textrm{36.0--145.0}} & 1,397 & Y & \cite{Barreau:1983ht}\\
    1987& Bates & 0.54 & 37.1 & \multicolumn{1}{l}{\:\:\textrm{N/A}} &   & \cite{O'Connell:1987ag}\\
    1987& Bates & 0.73 & 37.1 & 54 & Y & \cite{O'Connell:1987ag}\\
    1988& Yerev. & \multicolumn{1}{c}{\textrm{1.93, 2.13}} & \multicolumn{1}{c}{\textrm{16.0, 18.0}} & 134 & Y & \cite{Bagdasaryan:1988hp}\\
    1988& SLAC & 0.65 & \multicolumn{1}{c}{\textrm{33.0, 53.0}} & \multicolumn{1}{l}{\:\:\textrm{N/A}} &  & \cite{Baran:1988tw}\\
    1988& SLAC & \multicolumn{1}{c}{\textrm{1.30--1.65}} & \multicolumn{1}{c}{\textrm{11.95--13.54}} & 263 & Y & \cite{Baran:1988tw}\\
    1989& SLAC & \multicolumn{1}{c}{\textrm{0.96--1.50}} & 37.5 & 250 & Y & \cite{Sealock:1989nx}\\
    1993& SLAC & \multicolumn{1}{c}{\textrm{2.02--3.60}} & \multicolumn{1}{c}{\textrm{15.02--30.01}} & 316 & Y & \cite{Day:1993md}\\
    1994& SLAC & \multicolumn{1}{c}{\textrm{12.1--17.3}} & \multicolumn{1}{c}{\textrm{12.8--15.9}} & 7 & Y &\cite{Gomez:1993ri}\\
    1995& SLAC & \multicolumn{1}{c}{\textrm{2.02--5.12}} & \multicolumn{1}{c}{\textrm{35.51--56.64}} & 56 & Y & \cite{Arrington:1995hs}\\
    1998& JLab & 4.05 & \multicolumn{1}{c}{\textrm{15.0--74.0}} &  398 & Y & \cite{Arrington:1998hz,Arrington:1998ps}\\
    2010& JLab & 5.77 & \multicolumn{1}{c}{\textrm{18.0--50.0}} & 359 & Y & \cite{Fomin:2008iq,Fomin:2010ei}\\
    2018& JLab & 2.22 & 15.54 & 177 & Y & \cite{Dai:2018xhi}\\
    \end{tabular}
\end{ruledtabular}
\end{table}



We see from Fig.~\ref{fig:DUNE_wideRange_a} that the basic features seen in Fig.~\ref{fig:comparisons_C} are present across the range of kinematic parameters for $0.8 \lesssim\n q \lesssim 2.8$ GeV. Namely:
\begin{itemize}
\item[(i)] In the QE regime, the tails of the cross section extending to low energy transfers are underestimated by 100\%. As noted before, these tails are a clear evidence of the deficiency of the RFG model. Fortunately, they fall outside the shaded kinematic region of DUNE and thus are not the main concern for accurate Monte Carlo simulations of long-baseline experiments.
\item[(ii)] For the $\Delta$ excitation, the  trend is the same as in Figs.~\ref{fig:comparisons_C_b} and~\ref{fig:comparisons_C_c}. 
The crossing of the prediction and data at the tail of the $\Delta$ resonance is not limited to the kinematics of Fig.~\ref{fig:comparisons_C_c}, but turns out to be quite generic, as indicated by the pallid stripes extending between the dotted and dashed lines at $0.5\lesssim\n q\lesssim2.8$ GeV.
It results in the predictions differing from the data by up to $\pm10$\% through most of the region. In the $\Delta$ peak, the cross section is consistently underestimated: by $\sim$30\%--40\% at $\n q \sim 1$ GeV and $\sim$5\%--10\% at $\n q \sim 2$ GeV. We will return to these observations in Sec.~\ref{sect:diagnosis}. As noted before, in the resonance regime the cross sections are large, so that, even when the relative differences appear modest, the absolute differences are nonetheless significant.
\item[(iii)] In the DIS channel, \genie{} significantly overestimates the data, as we have seen before in Fig.~\ref{fig:comparisons_C}. We see that this region overlaps with the high event density for DUNE, and hence has high experimental significance. We will return to this issue in Sec.~\ref{sect:diagnosis}.
\end{itemize}

\begin{figure*}
\centering
    \includegraphics[width=0.90\textwidth]{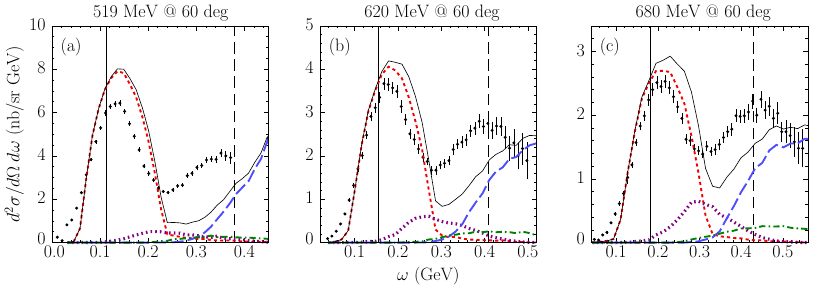}
\caption{\label{fig:Delta} Comparisons of the position of the quasielastic and $\Delta$ peaks in electron scattering off carbon according to the \genie{} predictions and the measurements of Barreau \etal{}~\cite{Barreau:1983ht}. For reference, the horizontal solid and dashed lines show the kinematics of quasielastic interaction and $\Delta$ excitation on free nucleons, respectively.}
\end{figure*}

Next, we examine Fig.~\ref{fig:DUNE_wideRange_b}, which expands the upper range of momentum transfer to 6 GeV. We see that the discrepancies in the DIS regime persist, in fact, they increase from 32\%
at 2.7 GeV to $\sim$36\% at 3.5 GeV, and reach 55\% at 5.3 GeV.

Moreover, large discrepancies appear in other regimes, even near the QE curve, where the data are overestimated by up to $\sim$65\% at $\n q\sim3.7$ GeV. Of course, in this kinematic regime, the QE feature cannot be clearly seen and DIS dominates. These findings confirm behavior observed in Fig.~\ref{fig:comparisons_C_a}.

Finally, another feature conspicuous in Fig.~\ref{fig:DUNE_wideRange} lies \emph{at lower momentum transfers}, $\n q \lesssim 0.8$ GeV, where we see another distinct region of large discrepancies. This second region falls into the regime of the second oscillation peak at DUNE. It is also relevant to the neutrino experiments performing measurements in the $\sim$1-GeV energy region, such as MicroBooNE~\cite{Acciarri:2016smi}, T2K~\cite{Abe:2011ks}, and Hyper-Kamiokande~\cite{Abe:2015zbg}.
Thus, the sub-GeV discrepancies deserve a closer look. Accordingly, we zoom in on this region in Fig.~\ref{fig:DUNE_narrowRange}.

Due to differences in their behavior, we present separately the results for scattering angles above and below $80^\circ$. At low scattering angles, the QE cross section is overestimated in the peak region by more then 100\% in 13 out of 42 datasets, and by more than 50\% in 26 datasets. For the QE peaks corresponding to $\n q\gtrsim0.5$ GeV, the agreement is visibly better, and \genie{} results typically differ from data by 10\%--20\%, see the red lines in the QE region turn more pallid in Fig.~\ref{fig:DUNE_narrowRange_a}.

At high scattering angles, the data are available from a single experiment performed at Saclay~\cite{Barreau:1983ht}, and do not extend beyond the onset of $\Delta$ excitation, limiting comparisons to the QE peak, see Fig.~\ref{fig:DUNE_narrowRange_b}. At this kinematics, the discrepancies between \genie{} and experimental results are particularly dire: they exceed 100\% for some points in the peak region in 16 out of 22 datasets, and are higher than 50\% for every dataset.

Of course, the RFG model---which describes the nucleus as a fragment of infinite noninteracting nuclear matter---cannot be expected to provide an accurate estimate of the cross sections at such low momentum transfers, where details of the shell structure are relevant and the effect of final-state interactions is sizable~\cite{Ankowski:2014yfa}.
However, the severity of the discrepancies we observe in our comparisons at large scattering angles, as well as the overestimated cross section at high momentum transfers, suggest that some implementation issues are at play~\cite{Ankowski:2014yfa,Butkevich:2005ph}. For example, the data are overestimated by up to $\sim$490\%  at $\n q\sim0.2$ GeV, which cannot be attributed to the limitations of the RFG model.

Next, let us turn to the dip region, between the QE and $\Delta$ peaks. The experimental cross sections here are underestimated by 50\%--80\% for $\n q\lesssim0.5$ GeV, but this discrepancy reduces to 10\%--25\% for the two datasets corresponding to $\n q\gtrsim0.8$ GeV. At the $\Delta$ peak, a similar trend can be observed, with the discrepancy decreasing from $\sim$60\% at $\n q\sim0.4$ GeV to $\sim$20\% at $\n q\sim0.8$ GeV. In the high-$\omega$ tail of the $\Delta$ peak, \genie{} results underestimate data by 10\%--15\% at $\n q\sim0.5$ GeV and by 5\%--10\% at $\n q\sim0.8$ GeV.

To see what happens here, in Fig.~\ref{fig:Delta} we explicitly show comparisons with three representative datasets. For reference, the vertical dashed (solid) line shows the position of the $\Delta$ (QE) peak in scattering on free nucleons. We immediately notice that the location of the $\Delta$ peak is systematically shifted to higher values of $\omega$, and that this shift is much larger than the one for the QE peak. Is the issue with the  $\Delta$ properties in \genie{}, or with the nuclear model?

Recall, that the same question arises in connection with the systemic discrepancies we observed in the DIS regime: does their origin lie in the hadronic physics of the primary vertex or in the nuclear model?

To understand better both of these issues, as well as other problems related to pion production, we are next going to analyze predictions of \genie{} in the case where they are not subject to significant nuclear effects.


\section{Isolating hadronic discrepancies with deuteron and hydrogen data}
\label{sect:diagnosis}

\begin{figure}
\centering
    \subfigure{\label{fig:comparisons_D_a}}
    \subfigure{\label{fig:comparisons_D_b}}
    \subfigure{\label{fig:comparisons_D_c}}
    \includegraphics[width=0.85\columnwidth]{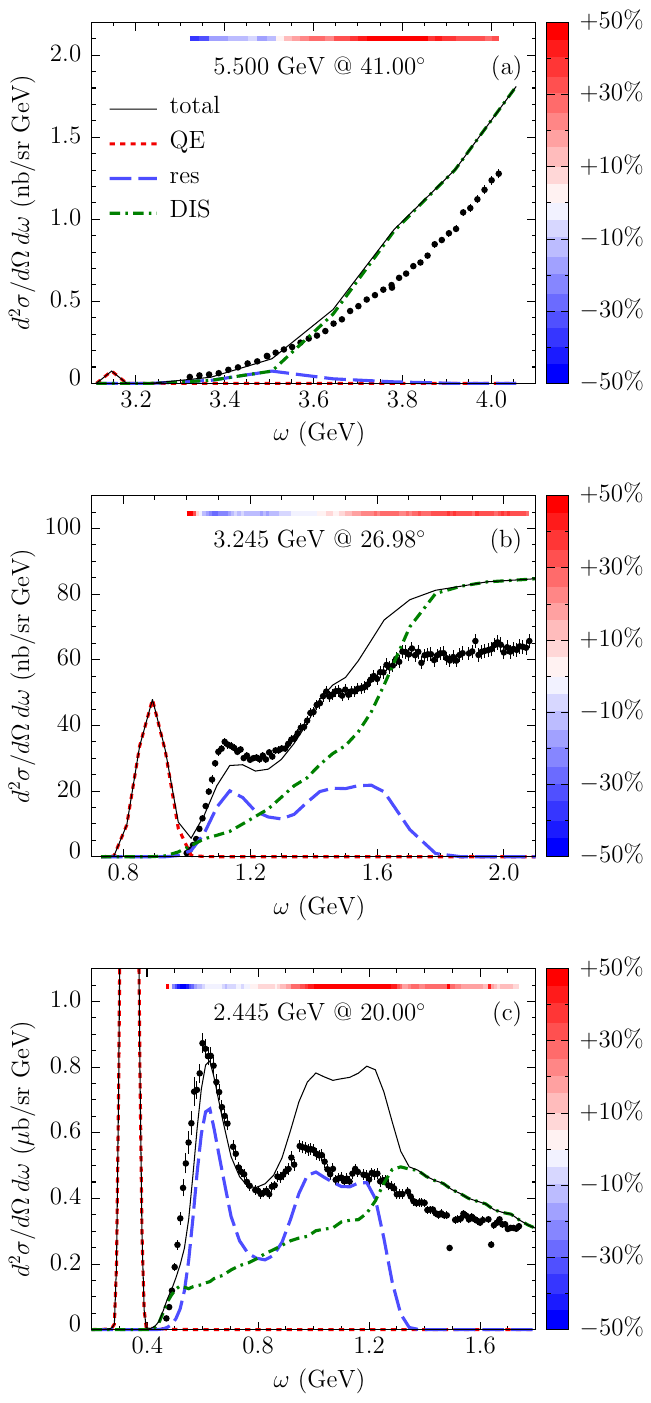}
\caption{\label{fig:comparisons_D} Same as Fig.~\ref{fig:comparisons_C} but for the double differential cross section for electron scattering off deuteron~\cite{Niculescu:1999,Niculescu:2000tk,Malace:2006,Malace:2009kw}, with the quasielastic contributions subtracted.}
\end{figure}

\begin{figure}
\centering
    \subfigure{\label{fig:comparisons_H_a}}
    \subfigure{\label{fig:comparisons_H_b}}
    \subfigure{\label{fig:comparisons_H_c}}
    \includegraphics[width=0.85\columnwidth]{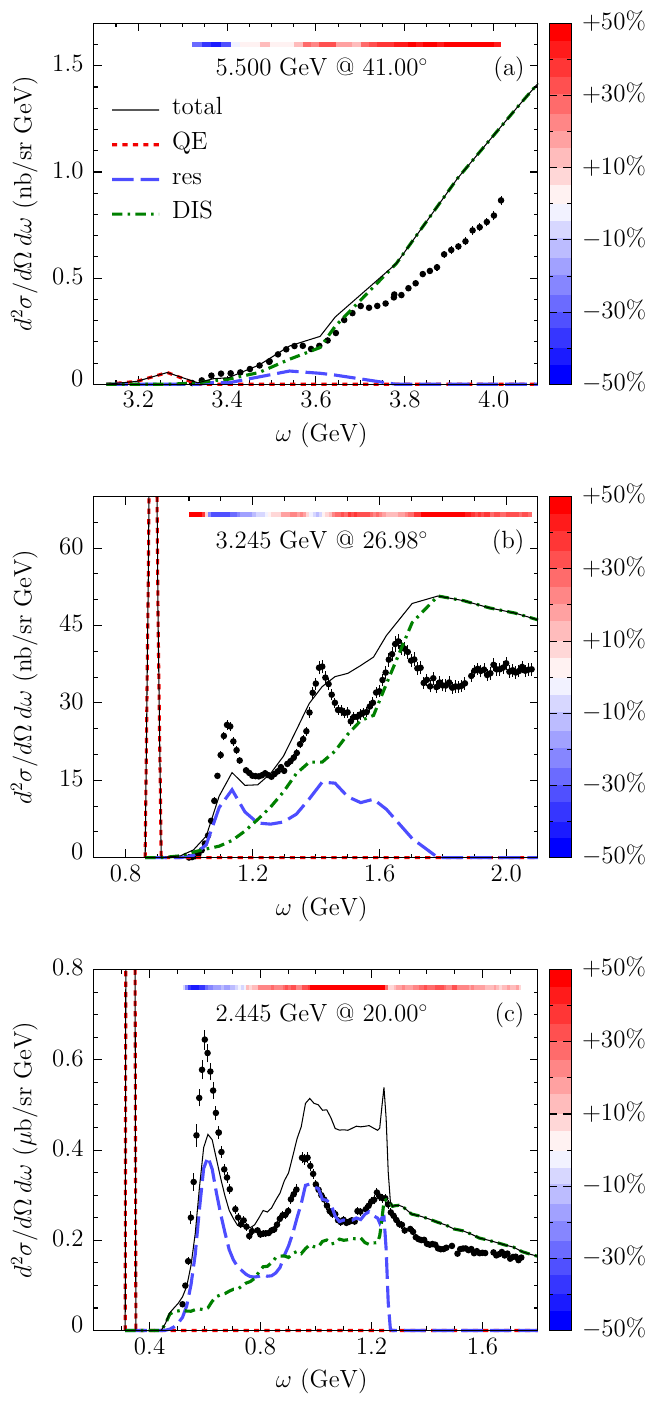}
\caption{\label{fig:comparisons_H} Same as Fig.~\ref{fig:comparisons_C} but for the double differential cross section for electron scattering off proton~\cite{Niculescu:1999,Niculescu:2000tk,Malace:2006,Malace:2009kw}, with the quasielastic contributions subtracted.}
\end{figure}

To this end, in Figs.~\ref{fig:comparisons_D} and~\ref{fig:comparisons_H}, we show comparisons of the \genie{} calculations with the data for electron scattering on deuteron and proton, reported in Refs.~\cite{Niculescu:1999,Niculescu:2000tk,Malace:2006,Malace:2009kw}.
Note that the quasielastic contributions are subtracted from these data.
The obtained results turn out to be consistent with the findings for carbon in Fig.~\ref{fig:comparisons_C}. Examining Fig.~\ref{fig:comparisons_D_a} [Fig.~\ref{fig:comparisons_H_a}] and Fig.~\ref{fig:comparisons_C_a}, corresponding to similar kinematics, one can observe for deuteron [proton] an excess of the predicted cross section in the DIS-dominated region, up to $\sim$53\% [$\sim$51\%], similar to that for carbon. Resonance excitation in the $\Delta$ region is underestimated, by $\sim$35\% [$\sim$40\%], in agreement with the result for carbon.

In Figs.~\ref{fig:comparisons_D_b} and~\ref{fig:comparisons_H_b}, the issues visible above the QE peak resemble those for carbon seen in Fig.~\ref{fig:comparisons_C_b}, corresponding to similar kinematics, with the cross section underestimated in the $\Delta$ region by up to $\sim$25\%--30\% for deuteron and $\sim$30\%--35\% for proton, and overestimated in the DIS-dominated region by $\sim$25\%--37\% for deuteron and by $\sim$26\%--51\% for proton.

The features of Figs.~\ref{fig:comparisons_D_c} and~\ref{fig:comparisons_H_c}, such as underestimated strength of the $\Delta$ peak and the excess of the predicted cross section in the region where higher resonances contribute, also bear a~close resemblance to the carbon results in Fig.~\ref{fig:comparisons_C_c}. For deuteron (proton) at this kinematics, the discrepancy amounts to
$\sim$43\% ($\sim$54\%) at $\omega=1$ GeV and reaches its maximum of $\sim$77\% ($\sim$86\%) at $\sim$1.2 GeV ($\sim$1.1 GeV). When energy transfer increases from 1.45 to 1.74 GeV, the discrepancy gradually decreases from $\sim$29\% ($\sim$34\%) to $\sim$6\% ($\sim$9\%). For comparison, the discrepancy with the carbon data in Fig.~\ref{fig:comparisons_C_c} increases to 47\% at $\omega=0.95$ GeV, at which energy transfer the measurement was stopped.

The consistency of the discrepancies observed for carbon, deuteron, and proton indicates that they originate from some issues related to the elementary cross sections, rather than from nuclear effects, which play a minimal role for deuteron and are completely absent for proton.

In an effort to unravel the role of the proton and neutron contributions in the observed discrepancies, Figs. \ref{fig:comparisons_D_c} and~\ref{fig:comparisons_H_c}  deliberately present results corresponding to the same kinematics. Their comparison suggests that above the $\Delta$ peak, most of the discrepancies for deuteron can be traced back to the proton cross section. In the $\Delta$ peak, the neutron cross section
seems to be too high, partly compensating the underestimated proton contribution, which leads to better agreement for deuteron than for proton.

Discussing the results presented in Figs.~\ref{fig:comparisons_C_c} and~\ref{fig:DUNE_wideRange} we have noticed that some carbon data collected at the tail of the $\Delta$ resonance are reproduced by \genie{} with accuracy exceeding 10\%. This behavior---absent for deuteron and proton---is likely to stem from Fermi motion of nucleons in carbon, which broadens all peaks. Redistributing part of the higher-resonance strengths to the $\Delta$ region, this effect reduces the cross section at high $\omega$ and increases it at low $\omega$, reducing two sources of discrepancy simultaneously. In view of the results of Figs.~\ref{fig:comparisons_D_c} and~\ref{fig:comparisons_H_c}, the improved agreement for carbon should be considered largely accidental.

\begin{table}
\caption{\label{tab:summary_D} Summary of the measurements of the cross section for inclusive electron scattering off deuteron. The datasets considered in this analysis are marked using ``Y''.
}
\begin{ruledtabular}
\begin{tabular}{@{}lldddcc@{}}
    \multirow{2}{*}{Year} & \multirow{2}{*}{Lab} & \multicolumn{1}{c}{\textrm{Energy}} & \multicolumn{1}{c}{\textrm{Angle}} & \multicolumn{1}{c}{Point} & \multicolumn{1}{c}{Incl.} & \multirow{2}{*}{Ref.}\\
  &  & \multicolumn{1}{c}{\textrm{(GeV)}} & \multicolumn{1}{c}{\textrm{(deg)}} & \multicolumn{1}{c}{number} & \multicolumn{1}{c}{here} & \\
    \hline
    1974& SLAC &  \multicolumn{1}{c}{\textrm{7.02--19.5}} & \multicolumn{1}{c}{\textrm{6.0, 10.0}} & 117 & Y &\cite{Poucher:1973rg,Whitlow:1991uw}\\
    1975& SLAC &  \multicolumn{1}{c}{\textrm{13.0--20.0}} & 4.0 & 31 & Y &\cite{Stein:1975yy,Whitlow:1991uw}\\
    1976& SLAC &  \multicolumn{1}{c}{\textrm{6.50--19.5}} & \multicolumn{1}{c}{\textrm{50.0, 60.0}} & 71 & Y &\cite{Atwood:1976ys,Whitlow:1991uw}\\
    1977& SLAC & \multicolumn{1}{c}{\textrm{6.52--18.4}} & 8.0 & 179 &  &\cite{Schutz:1976he}\\
    1979& SLAC &  \multicolumn{1}{c}{\textrm{4.50--20.0}} & \multicolumn{1}{c}{\textrm{18.0--34.0}} & 302 & Y &\cite{Bodek:1979rx,Whitlow:1991uw}\\
    1983& SLAC & \multicolumn{1}{c}{\textrm{6.50--19.5}} & \multicolumn{1}{c}{\textrm{6.0--20.6}} & 100 & Y &\cite{Mestayer:1982ba,Whitlow:1991uw}\\
    1986& Bates & \multicolumn{1}{c}{\textrm{0.22--0.32}} & 180.0 & 235 &  &\cite{Parker:1986}\\
    1988& Bates & \multicolumn{1}{c}{\textrm{0.17--0.60}} & \multicolumn{1}{c}{\textrm{60.0--134.5}} & \multicolumn{1}{l}{\:\:\textrm{N/A}} &  &\cite{Quinn:1988ua,Dytman:1988fi}\\
    1988& SLAC & \multicolumn{1}{c}{\textrm{0.84--1.28}} & 180.0 & 227 &  &\cite{Arnold:1988us}\\
    1990& SLAC & \multicolumn{1}{c}{\textrm{3.75--24.5}} &\multicolumn{1}{c}{\textrm{11.1--46.2}} & 70 &  &\cite{Whitlow:1991uw,Gomez:1993ri}\\
    1992& SLAC & \multicolumn{1}{c}{\textrm{9.74--21.0}} & 10.0 & 425 &  &\cite{Rock:1991jy}\\
    1992& SLAC & \multicolumn{1}{c}{\textrm{1.51--2.84}} & \multicolumn{1}{c}{\textrm{41.1--90.1}} & 179& Y &\cite{Lung-thesis:1992,Lung:1992bu}\\
    1994& SLAC & \multicolumn{1}{c}{\textrm{8.00--24.5}} & \multicolumn{1}{c}{\textrm{11.1--22.2}} & 23 & Y &\cite{Gomez:1993ri}\\
    1996& SLAC & \multicolumn{1}{c}{\textrm{2.02--5.12}} & \multicolumn{1}{c}{\textrm{38.8--56.6}} & 56 &  &\cite{Arrington:1995hs}\\
    1998& SLAC & 5.51 & \multicolumn{1}{c}{\textrm{15.1--26.8}} & 188& Y &\cite{Stuart:1996zs}\\
    1999& JLab & 4.05 & \multicolumn{1}{c}{\textrm{15.0--55.0}} &  386 & Y &\cite{Arrington:1998hz,Arrington:1998ps}\\
    2000& JLab & \multicolumn{1}{c}{\textrm{2.45--4.05}} & \multicolumn{1}{c}{\textrm{20.0--70.0}} &  699 & Y &\cite{Niculescu:1999,Niculescu:2000tk}\\
    2009& JLab & 5.50 & \multicolumn{1}{c}{\textrm{37.9--70.0}} &  261 & Y &\cite{Malace:2006,Malace:2009kw}\\
    2010& JLab & 5.77 & \multicolumn{1}{c}{\textrm{18.0--50.0}} & 260 & Y &\cite{Fomin:2008iq,Fomin:2010ei}\\
    \end{tabular}
\end{ruledtabular}
\end{table}



\begin{table}
\caption{\label{tab:summary_H} Summary of the measurements of the cross section for inclusive electron scattering off proton. The datasets considered in this analysis are marked using ``Y''.
}
\begin{ruledtabular}
\begin{tabular}{@{}lldddcc@{}}
    \multirow{2}{*}{Year} & \multirow{2}{*}{Lab} & \multicolumn{1}{c}{\textrm{Energy}} & \multicolumn{1}{c}{\textrm{Angle}} & \multicolumn{1}{c}{Point} & \multicolumn{1}{c}{Incl.} & \multirow{2}{*}{Ref.}\\
  &  & \multicolumn{1}{c}{\textrm{(GeV)}} & \multicolumn{1}{c}{\textrm{(deg)}} & \multicolumn{1}{c}{number} & \multicolumn{1}{c}{here} & \\
    \hline
    1974& SLAC & \multicolumn{1}{c}{\textrm{7.02--19.5}} & \multicolumn{1}{c}{\textrm{6.0, 10.0}} &  117 & Y &\cite{Poucher:1973rg,Whitlow:1991uw}\\
    1975& SLAC & \multicolumn{1}{c}{\textrm{13.0--20.0}} & 4.0 &  32 & Y &\cite{Stein:1975yy,Whitlow:1991uw}\\
    1976& SLAC & \multicolumn{1}{c}{\textrm{6.50--19.5}} & \multicolumn{1}{c}{\textrm{50.0, 60.0}} &  77 & Y &\cite{Atwood:1976ys,Whitlow:1991uw}\\
    1979& SLAC & \multicolumn{1}{c}{\textrm{4.50--20.0}} & \multicolumn{1}{c}{\textrm{15.0--34.0}} &  316 & Y &\cite{Bodek:1979rx,Whitlow:1991uw}\\
    1983& SLAC & \multicolumn{1}{c}{\textrm{6.50--20.0}} & \multicolumn{1}{c}{\textrm{6.0--20.6}} &  119 & Y &\cite{Mestayer:1982ba,Whitlow:1991uw}\\
    1998& SLAC & \multicolumn{1}{c}{5.51, 9.80} & \multicolumn{1}{c}{\textrm{13.2--26.8}} & 113& Y &\cite{Stuart:1996zs}\\
    2000& JLab & \multicolumn{1}{c}{\textrm{2.45--4.05}} & \multicolumn{1}{c}{\textrm{20.0--70.0}} &  742& Y &\cite{Niculescu:1999,Niculescu:2000tk}\\
    2004& JLab & \multicolumn{1}{c}{\textrm{1.15--5.50}} & \multicolumn{1}{c}{\textrm{12.5--78.0}} &  1,273 & Y &\cite{Liang:2003,Liang:2004tj}\\
    2009& JLab & 5.50 & \multicolumn{1}{c}{\textrm{37.9--70.0}} &  261 & Y &\cite{Malace:2006,Malace:2009kw}\\
    %
    \end{tabular}
\end{ruledtabular}
\end{table}

\begin{figure}
\centering
    \includegraphics[width=0.85\columnwidth]{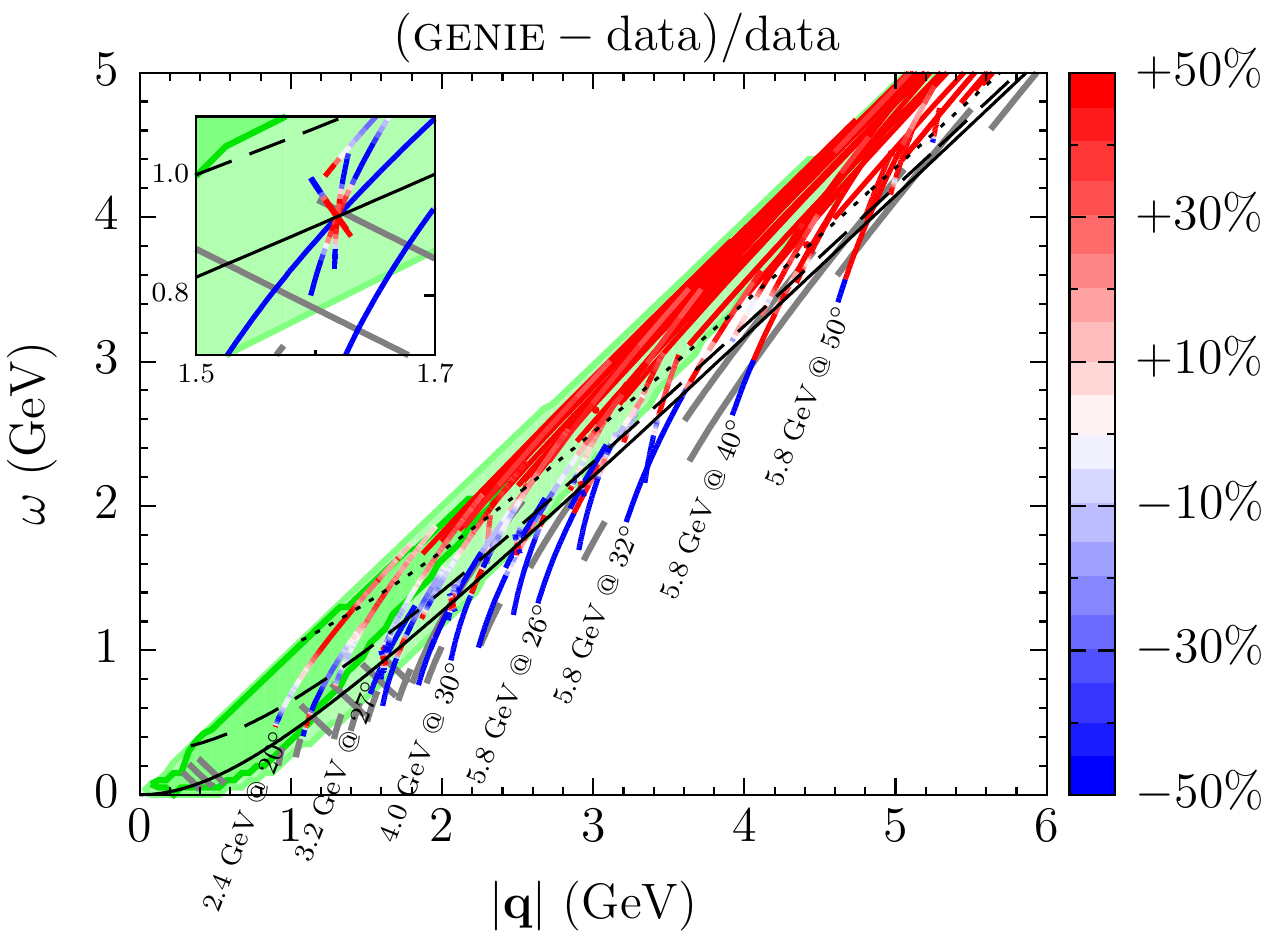}
\caption{\label{fig:DUNE_deuteron} Same as Fig.~\ref{fig:DUNE_wideRange} but for deuteron. The thick grey lines represent the datasets not included in this analysis. The inset magnifies one of the regions probed at different kinematic setups.}
\end{figure}

The global picture emerging from the deuteron data listed in Table~\ref{tab:summary_D} is shown in Fig.~\ref{fig:DUNE_deuteron}. Note that the thick grey lines, such as those in the bottom left corner, show the datasets not included in our analysis, limited to the vicinity of the QE peak. In the DIS region, the deuteron measurements cover the DUNE kinematics way better then the carbon ones, and provide clear evidence that \genie{} consistently overpredicts the cross section to a significant extent. At momentum transfer 1.5 GeV this discrepancy amounts to 20\%--30\%, increasing to 40\%--50\% at 2.4 GeV, and to 60\%--70\% at 4.5 GeV.

What is the origin of such large discrepancies? As described in Sec.~ \ref{sect:gutsofgenie}, \genie{} treats the DIS regime using the approach of Bodek and Yang~\cite{Bodek:2002ps,Bodek:2004pc}. Yet, this phenomenological model is constructed to give a good fit to a large body of DIS data available for deuteron, \emph{including} those of Refs.~\cite{Poucher:1973rg,Stein:1975yy,Atwood:1976ys,Bodek:1979rx,Schutz:1976he,Whitlow:1991uw,Mestayer:1982ba,Gomez:1993ri} used in Fig.~\ref{fig:DUNE_deuteron}. This makes the discrepancies revealed in this figure especially confounding and suggests that the implementation of the approach of Bodek and Yang in \genie{} should be scrutinized.

In particular, Figs.~\ref{fig:comparisons_D} and \ref{fig:comparisons_H} suggest that the way the DIS component is combined with the contribution of the higher resonances results in double-counting below $\omega=1.2$ GeV. Notice that the approach of Bodek and Yang is designed to include all resonance contributions, with the exception of $\Delta$(1232), see Ref.~\cite{Bodek:2004pc}. Another, known, limitation of the \genie{} implementation of resonance production is that the interference effects from the original model of Rein and Seghal  are neglected.

In contrast with the higher resonances, the \genie{} predictions for the $\Delta$ resonance are systematically underestimated, typically by 20\%--40\%, particularly at the kinematics corresponding to low $\n q$ and low $\omega$. The difference between \genie{} and data decreases when the momentum transfer increases. For example, at momentum transfer 1.8 GeV it amounts to 30\%--40\%, but at 3.0 GeV it reduces to 20\%--30\%.

In nearly all analyzed cases, the QE peak is overestimated. The inset of Fig.~\ref{fig:DUNE_deuteron} shows that, qualitatively, a consistent picture emerges from data probing the QE peak at different kinematics, but the discrepancy increases for high scattering angles. For example, the \genie{} results reproduce with $\sim$40\% accuracy the height of the QE peak in the data for 5.51 GeV and $15^\circ$, 2.41 GeV and $41^\circ$, and 1.97 GeV and $55^\circ$. However, the discrepancy reaches $\sim$70\% for 1.51 GeV and $90^\circ$. In every analyzed dataset for deuteron, the low-$\omega$ tail of the QE peak is underestimated, by up to 100\%.

\begin{figure}
\centering
    \includegraphics[width=0.85\columnwidth]{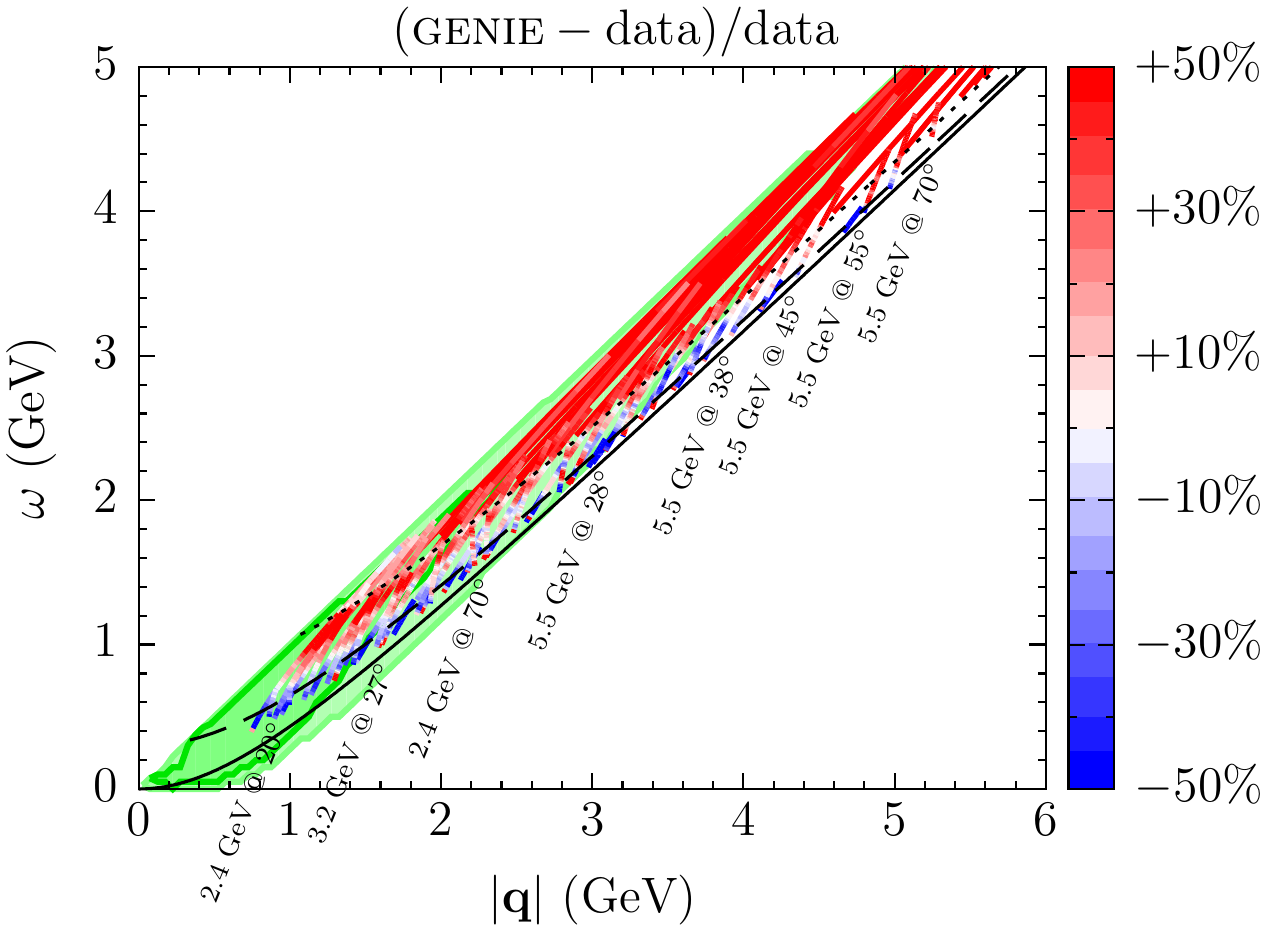}
\caption{\label{fig:DUNE_proton} Same as Fig.~\ref{fig:DUNE_wideRange} but for proton.}
\end{figure}

As deuteron data are not completely void of nuclear effects, in Fig.~\ref{fig:DUNE_proton} we present a comparison of \genie{} predictions with the data for inclusive electron scattering on proton summarized in Table~\ref{tab:summary_H}. One can observe a consistent pattern emerging from data collected at different kinematics, which shows that the results depicted in Fig.~\ref{fig:comparisons_H} can be considered a representative sample. In the DIS regime---extending above the dotted line---the electron-proton cross section from \genie{} is overestimated, and the discrepancy exhibits an increasing trend when momentum transfer increases: amounting to 25\%--35\% at 1.5 GeV, it rises to 35\%--45\% at 3.2 GeV, and to 45\%--55\% at 5.1 GeV.

The proton results in the region of pure DIS confirm our findings for deuteron. Also in this case, unaffected by nuclear effects, the data that the approach of Bodek and Yang describes by construction~\cite{Poucher:1973rg,Stein:1975yy,Atwood:1976ys,Bodek:1979rx,Mestayer:1982ba,Whitlow:1991uw} are overestimated by its implementation in \genie{} by as much as 60\%--110\%.


At the kinematics where higher resonances can be excited---delimited by the dotted and dashed lines corresponding to the onset of DIS and the $\Delta$ peak, re{\-}spec{\-}tive{\-}ly---the discrepancy forms a complicated pattern, illustrating that these mechanism of interaction are not accurately accounted for in \genie{}, and the positions and widths of the peaks in the data differ from the predictions. Pion production on protons in the $\Delta$ resonance region turns out to be underestimated in \genie{} typically by 20\%--40\% over the whole considered kinematics.

The results for deuteron and proton clearly show that the main source of issues related to pion production in \genie{} observed for carbon and heavier nuclei lies not in nuclear effects, but rather in the description of the elementary cross sections for scattering on both protons and neutrons. As a concrete example, we observe that \genie{} can reproduce with fairly good accuracy the carbon to deuteron cross-section ratios reported by Gomez \etal~\cite{Gomez:1993ri} for several points at $Q^2=5$ GeV$^2$, although individually these cross sections deviate from the data by 60\%--95\%.

\section{Discussion}
\label{sect:discussion}

When testing a computer model against data, finding disagreements is only the first stage of the process. The next, key, question is whether one can identify concrete physical processes that are mismodeled. Finally, the best outcome would be to identify specific, constructive improvements. While the full program of this type is beyond the scope of any one paper, let us organize our findings with these points in mind.

{\it Quasielastic scattering}. Examining our heat-map plots, we saw that a certain amount of discrepancy is clearly present in this regime at low energy transfers. Even though such discrepancies may appear large on these plots, their impact should nevertheless not be overinterpreted. First of all, these large percentage discrepancies often arise in the regions where the absolute rates are small. Second, the agreement can be improved by a subtle shift of the peak and other adjustments~\cite{Bodek:2020wbk}, which are well understood theoretically~\cite{Ankowski:2014yfa}.

Below momentum transfer values $\n q \sim0.8$ GeV, and above $\n q \sim3$ GeV, the QE electron scattering rate predicted by \genie{} does show significant discrepancies from the data. As we discussed, the severity of these discrepancies makes it likely that---in addition to the theoretically understood corrections to the RFG model---an implementation issue is likely at play.

Above all, what is noteworthy in our comparison is just how well the simple RFG model, as currently implemented in GENIE, produces a reasonably good description of the quasielastic peak at energies of 1--3 GeV. This agreement of the model is nontrivial and very significant, as it imposes important constraints on how much MEC component can be added to the model. An excessively large MEC contribution would distort the right side of the QE peak, spoiling the agreement. We will return to this below.

{\it $\Delta$ resonance production.} The onset of inelasticity and the $\Delta$ resonance offers example where not only problems, but also a path to improvement, can be identified. The location of the $\Delta$ resonance in carbon and other complex nuclei is found to occur at systematically higher energy transfer values than required by data. At the same time, it is gratifying that the $\Delta$ peak is in the correct place for hydrogen: this indicates that the underlying hadronic physics is correct and one should reexamine the implementation of $\Delta$ production within the nuclear framework.
Fixing the location of $\Delta$ peak in complex nuclei would have two further important implications:
\begin{itemize}
\item[(i)] Shifting the $\Delta$ peak closer to the QE peak would also make the MEC component in the dip more manageable, making it easier to explain the data without distorting the shape of the QE peak, as mentioned above.
\item[(ii)] The benefits would go beyond achieving better agree{\-}ment with inclusive data. MEC and pion pro{\-}duc{\-}tion predict different composition of the final hadronic system. Correctly modeling the fractions of each process is very important for several neutrino experiments, as discussed below.
\end{itemize}

{\it Higher resonances and deep inelastic regime.} The large discrepancies revealed by our study in the region of higher resonances are perhaps the most surprising result of our analysis. Unlike the $\Delta$ peak case, these discrepancies are traced to the underlying hadronic physics, by comparing the generator output to deuterium and hydrogen data. The nature of these discrepancies thus requires detailed studies of hadron production in the so-called ``shallow inelastic regime''.
\begin{itemize}
\item[(i)] This comparison, yet again, illustrates the power of electron scattering: the discrepancies become less pronounced when integrated over a range of electron energies corresponding to the width of the DUNE neutrino beam. The integration, however, only masks the problem and may lead to systematic misconstruction of neutrino energy in DUNE. It is very desirable to understand if the hadronic discrepancies in this regime can already be seen in available neutrino data, e.g., at NOvA, MicroBooNE/ICARUS, and MINERvA. Composition of the hadronic system provides powerful information about the interaction physics.
\item[(ii)] It is noteworthy that a consistent implementation of the Bodek-Yang approach should show significantly better agreement with data, at the inclusive level. Thus, one should examine its current implementation in GENIE. In particular, it appears from the analysis of the hydrogen and deuterium data [cf. Figs.~\ref{fig:comparisons_D_c} and \ref{fig:comparisons_H_c}] that the implementation in GENIE may be double-counting the contributions from the resonant and deep-inelastic components in this regime where both are present.
\item[(iii)] In any case, a more accurate modeling of hadronic effects in this regime is necessary. This requires a combined theory-experiment effort, with theory frameworks that respect quark-hadron duality benchmarked against new experimental data with detailed information about hadronic final states. An example of an experiment that could accomplish this is provided by CLAS and by the LDMX setup~\cite{Ankowski:2019mfd}.
\end{itemize}

\begin{figure}
\centering
    \includegraphics[width=0.85\columnwidth]{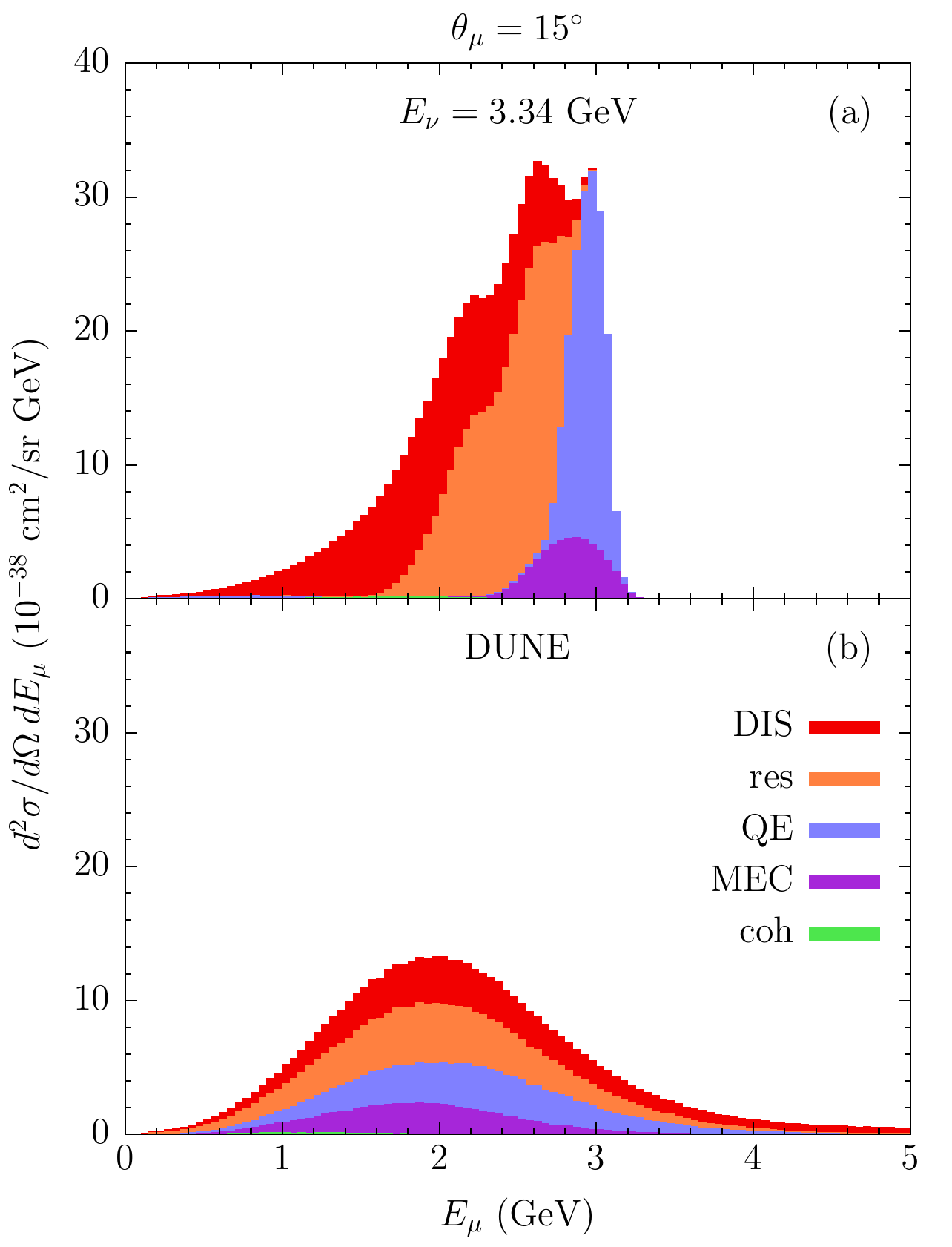}
\caption{\label{fig:flux_integration} Double differential cross section for charged-current scattering of muon neutrinos on carbon at $15^\circ$, obtained using \genie{} for (a) the fixed beam energy of 3.34 GeV, the average DUNE energy, and (b) the near-detector flux of DUNE~\cite{DUNE_flux}, shown as stacked histograms.}
\end{figure}

\begin{figure}
\centering
    \includegraphics[width=0.85\columnwidth]{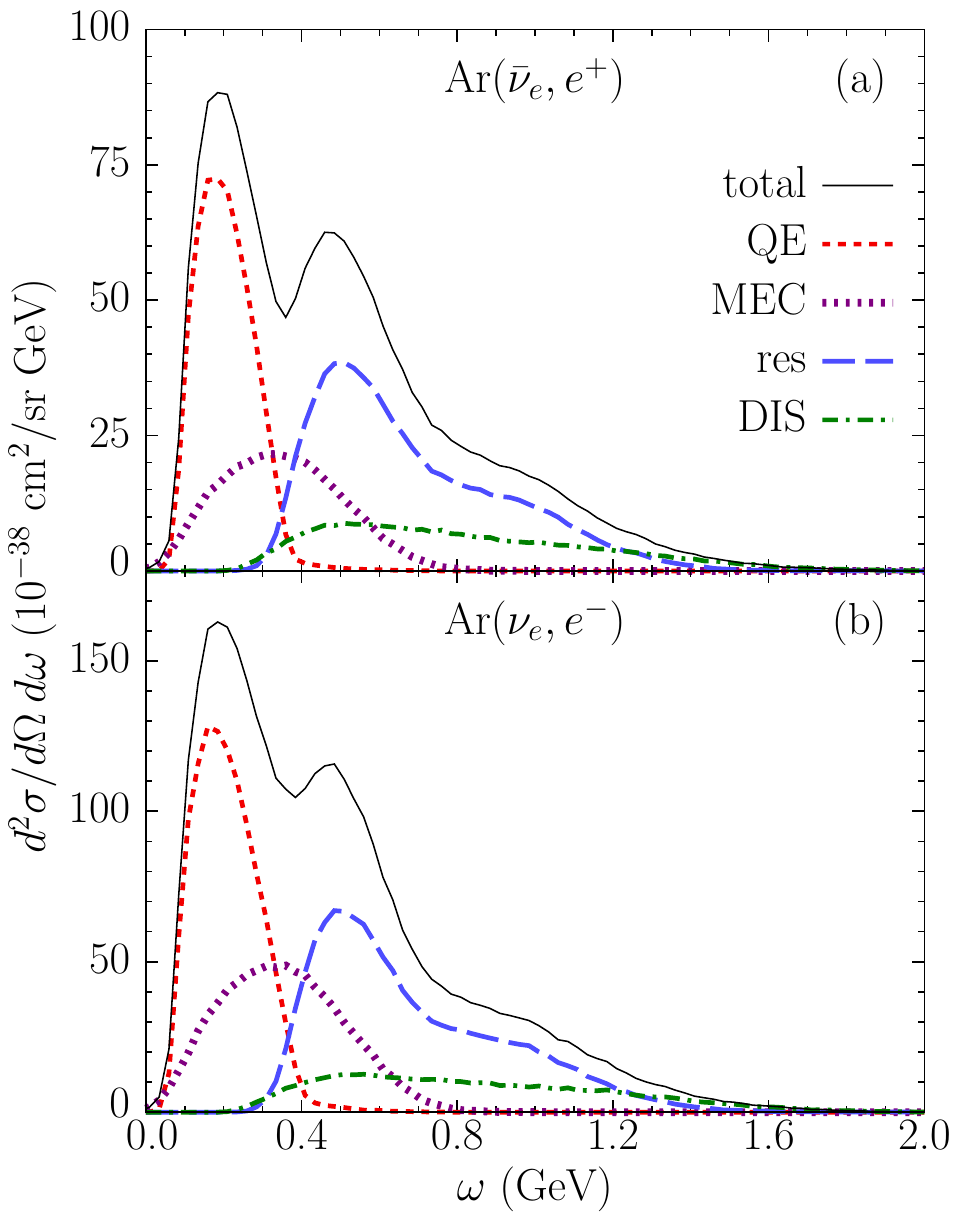}
\caption{\label{fig:Ar_nu&anu} Cross sections for 2.222-GeV electron (a) antineutrino and (b) neutrino scattering off argon obtained for scattering angle $15.54^\circ$ using \genie{}.}
\end{figure}

Finally, a connection to neutrino data should be discussed.
As we have seen here, electron scattering affords tight control over kinematics, allowing one to isolate specific physical processes and consider multiply differential cross sections. This makes them invaluable for testing specific physics ingredients in generator codes.

The power of knowing the exact kinematics is illustrated graphically in Fig.~\ref{fig:flux_integration}. In the top panel, we consider a hypothetical situation with a monochromatic neutrino beam and a fixed scattering angle. We see that the different scattering mechanisms separate rather clearly according to the value of the produced muon's energy. In the bottom panel, we perform the same exercise for the actual energy spectrum of the DUNE beam. The separation of the different contributions is clearly lost.

Neutrino scattering, on the other hand, offers measurements of axial effects. Modern neutrino detectors, particularly liquid-argon detectors, also offer a high degree of containment and particle-identification capabilities, potentially allowing a detailed study of the final-state hadronic system. At the same time, for calorimetric energy reconstruction, the same experiments \emph{rely} on generators to predict the properties of the hadronic system. These properties include the composition (fractions of pions/protons/neutrons) and subthreshold charge clusters. Thus, a combined analysis of electron and neutrino scattering experiments should offer highly complementary benefits.

To this end, for the electron-scattering comparison to be a powerful tool, the generator must implement \emph{the same} physics in electron and neutrino modes. {\it Ad hoc} adjustments implemented in the neutrino channel, even to improve phenomenological agreement with data, break the powerful link between the two probes.
Fortunately, \genie{} is built to simulate the nuclear scattering of neutrinos and  electrons in the same physics framework and this consistency has been emphasized in developing version 3 of the code~\cite{Roda:2017ovf}.

As an example, consider Fig.~\ref{fig:Ar_nu&anu}, which depicts a neutrino analog of the electron-scattering data from Fig.~\ref{fig:CTi}. Namely, the cross sections for 2.2-GeV $\bar\nu_e$ and $\nu_e$ scattering off argon at $15.54^\circ$, obtained using \genie{}. As we noted before, in electron scattering an addition of an excessively large MEC component creates a tension with the measured shape of the quasielastic peak. In the case of neutrinos, the default setting of the generator multiples the normalization of this component by the factor of 9, compared with electrons. Such {\it ad hoc} modifications are not desirable and, in any case, need a detailed investigation.

\section{Conclusions}
\label{sect:conclusions}

We have seen that electron-scattering data can serve as a very effective tool for testing event generators. This is made possible by two principal factors. First, it features precisely known electron kinematics: initial and final energies, as well as the scattering angle. This allows one to individually examine different physical processes, from quasielastic to deep inelastic scattering. Second, there are numerous electron-scattering datasets covering the range of energies and angles relevant to DUNE, and spanning different interaction regimes.

It is worth noting that at the considered level of details, electron-scattering data turn out to be remarkably consistent between different experiments, kinematic settings, and targets. Because these data have not been previously used to tune the parameters of the GENIE generator, they present a great opportunity to test the cross-section estimates, identify paths towards their improvement, and provide means of determining systematic uncertainties in neutrino-oscillation experiments.

To broadly summarize our findings, we have observed persistent disagreements between the GENIE predictions and electron-scattering data in the few-GeV energy range. This range spans multiple physical phenomena and our comparisons reveal significant tensions across these scattering regimes, with most significant---and surprising---discrepancies found in the region of large inelasticity.

The full utility of our analysis comes from its ability to identify different constructive strategies that, going forward, can improve the situation. As an illustration, an immediate reduction of the normalization of the DIS component by 28\% would improve agreement with the electron scattering data across a slice of the DUNE-relevant phase space, as we show in Appendix B. We also demonstrate there the limitations of such a phenomenological tuning approach. It may be used as a temporary measure, but not as a substitute for implementing the correct functional form of the approach of Bodek and Yang~\cite{Bodek:2002ps,Bodek:2004pc}.


It is worthwhile to briefly return to the starting point for our discussion, provided by the comparisons in Fig.~\ref{fig:Barreau}. We have now seen that the systematic exploration of the data reveals issues not only with the MEC contribution, but also with the implementations of the QE scattering and the $\Delta$-resonance excitation. Moreover, because of the limitations of the datasets used, the comparison entirely missed the large discrepancies in the region of higher resonances and DIS. This illustrates the importance of taking a global view, across all scattering regimes, which allows one to identify problems across the ranges. Moreover, by correcting physics in one regime we can improve the treatment of another. As an explicit example, fixing the position of the $\Delta$ peak and taking the accurate form of the QE peak into account influences how one treats the MEC contribution.

We urge all experimental groups performing electron-scattering studies, the CLAS Collaboration in particular, to publish the cross sections in tabulated form. As nuclear and hadronic effects are far from being fully understood, one can anticipate the nuclear and hadronic physics communities to find new applications for these cross sections for many years to come, likely for studies very different from the objective they were extracted for. In the short term, these data have the potential to stimulate improving accuracy of Monte Carlo generators employed in neutrino physics. In particular, the CLAS data has an excellent coverage in the kinematic space of DUNE, as summarized in Appendix~\ref{appendix:CLASdata}. We also strongly encourage future dedicated data-taking campaigns, at various experiments at Jefferson Laboratory and at the proposed LDMX experiment at SLAC~\cite{Ankowski:2019mfd}.

Finally, targeted analyses of all available neutrino data would also very desirable. In particular, the authors are intrigued whether the MINERvA experiment can observe neutrino-scattering counterparts to the DIS discrepancies revealed in the electron-scattering analysis of this paper. Studies of DIS were  prominently identified in the MINERvA science program from the start~\cite{Drakoulakos:2004gn} and we strongly encourage the Collaboration to carry them out as part of the outgoing analysis campaign. Likewise, detailed measurements of the hadronic final states at MicroBooNE and ICARUS could shine light on the physics in the quasielastic, dip, and $\Delta$ regions.

\begin{acknowledgments}
We are very grateful to Steven Dytman for assistance with using GENIE in electron-scattering mode. We thank Adi Ashkenazi, Minerba Bettancourt, and Afroditi Papadopoulou for stimulating feedback, Alessandro Lovato and Noemi Rocco for discussions that helped to sharpen the presentation, and Raquel Fernandez Castillo for her support and encouragement. Our presentation also benefited from the feedback of John Arrington. This work is supported by the U.S. Department of Energy under Award No. DE-AC02-76SF00515.
\end{acknowledgments}



\appendix
\section{Review of the $(e,e')$ scattering data used in this study}
\label{appendix:data}

{\it Carbon. }Presenting results for medium-size nuclei, in this paper we focus on carbon, because this target is by far the most extensively studied in electron-scattering experiments~\cite{Benhar:2006er}. To ensure that the picture emerging from our analysis is complete, we include all available data~\cite{Whitney:1974hr,Barreau:1983ht,O'Connell:1987ag,Bagdasaryan:1988hp,Baran:1988tw,Sealock:1989nx,Day:1993md,Gomez:1993ri,Arrington:1995hs,Arrington:1998hz,Arrington:1998ps,Fomin:2008iq,Fomin:2010ei,Dai:2018xhi}, from measurements performed over a range of beam energies and scattering angles.

\begin{figure}[t]
\centering
    \subfigure{\label{fig:qW_a}}
    \subfigure{\label{fig:qW_b}}
    \subfigure{\label{fig:qW_c}}
    \includegraphics[width=0.90\columnwidth]{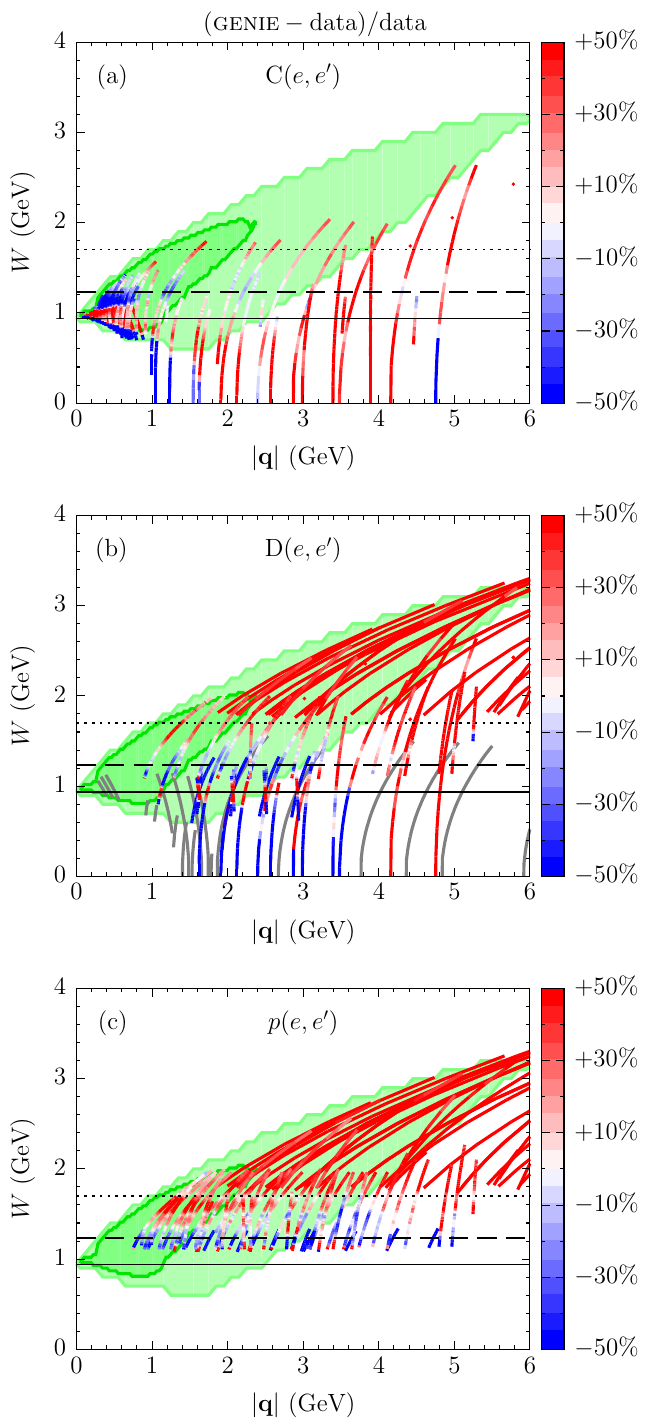}
\caption{\label{fig:qW} Fractional difference between the \genie{} calculations and the experimental cross sections for electron scattering off (a) carbon, (b) deuteron, and (c) proton presented on the plane of momentum transfer $\n q$ and hadronic mass $W$. The region corresponding to 68\% (95\%) of charged-current events in DUNE is shown as the dark (light) shaded area as in Fig.~\ref{fig:DUNE_vs_NOvA}.}
\end{figure}

In total, our analysis includes 3,446 data points for carbon, summarized in Table~\ref{tab:summary_C}. As Fig.~\ref{fig:qW_a} shows in the $(\n q, W)$ plane, the probed kinematic regimes range from quasielastic to deep-inelastic scattering. It is important to note, however, that the bulk of the measurements for carbon focused on the region of the quasielastic peak, studying short-range interactions between nucleons in nuclear medium. As a consequence, a large swath of the kinematic region of interest for long-baseline neutrino experiments remains unprobed by electron scattering, even in the best studied case of carbon. Below in subsection~\ref{Subsec:carbon} of this Appendix we review the literature reporting existing experimental results for carbon.

{\it Deuteron. } This target gives an excellent opportunity to test the accuracy of the elementary cross sections for scattering on neutrons and protons---used in the calculations for complex nuclei---in the case in which they are minimally affected by nuclear effects.
As the simplest nuclear system, deuteron was extensively studied in the past electron-scattering experiments~\cite{Resonance_database}, see the summary in Table~\ref{tab:summary_D} and Fig.~\ref{fig:qW_b}. Yet many cross-section measurements probed the kinematics corresponding to the QE peak and its  vicinity~\cite{Schutz:1976he,Parker:1986,Quinn:1988ua,Dytman:1988fi,Arnold:1988us,Rock:1991jy,Lung-thesis:1992,Lung:1992bu,Arrington:1995hs,Stuart:1996zs}. Vivid interest in deep-inelastic scattering brought a series of measurements, most of which were performed between mid-1970s to mid-1980s~\cite{Poucher:1973rg,Stein:1975yy,Atwood:1976ys,Bodek:1979rx,Mestayer:1982ba,Whitlow:1990gk,Whitlow:1991uw,Gomez:1993ri}, but beam energies in these experiments typically exceeded 8 GeV.
As a consequence, only the measurements~\cite{Arrington:1998hz,Arrington:1998ps,Niculescu:1999,Niculescu:2000tk,Malace:2006,Malace:2009kw,Fomin:2008iq,Fomin:2010ei} directly explored the region most relevant to DUNE.

In addition to the datasets covering resonance production and the DIS regime, we include as a representative sample the QE data of Refs.~\cite{Lung-thesis:1992,Lung:1992bu}---spanning a broad range of energies and scattering angles---to verify that they all lead to a consistent picture. In total, our analysis for deuteron is based on 2,617 data points, leaving 1,192 data points for future considerations, see Table~\ref{tab:summary_D}. We review the literature reporting the cross sections for deuteron in subsection~\ref{Subsec:deuteron} of this Appendix.
References~\cite{Gomez:1993ri,Arrington:1995hs,Arrington:1998hz,Arrington:1998ps,Fomin:2008iq,Fomin:2010ei} containing both deuteron and carbon data are discussed in subsection~\ref{Subsec:carbon}.

{\it Proton.} While analysis of deuteron data gives insight into the proton and neutron contributions combined, an even deeper level of understanding can be reached when comparisons to data for electron scattering on deuteron and protons can be made at the same kinematics. Proton data are also interesting in their own right, allowing the calculated cross sections to be tested in the case completely void of nuclear effects. For the proton target, we analyze 3,050 data points, reported in Refs.~\cite{Poucher:1973rg,Stein:1975yy,Atwood:1976ys,Bodek:1979rx,Mestayer:1982ba,Whitlow:1991uw,Stuart:1996zs,Niculescu:1999,Niculescu:2000tk,Malace:2006,Malace:2009kw,Liang:2003,Liang:2004tj}. The kinematics of these measurements is summarized in Table~\ref{tab:summary_H} and presented in Fig.~\ref{fig:qW_c}. References~\cite{Liang:2003,Liang:2004tj}, the only ones reporting proton measurements unaccompanied by the deuteron results, are discussed in subsection~\ref{Subsec:proton}.

\subsection{Carbon literature}
\label{Subsec:carbon}

Whitney \etal~\cite{Whitney:1974hr} reported for several targets ranging from lithium to lead--- including carbon---a measurement of the cross sections in the region of the QE peak at a single kinematic setup, performed in High Energy Physics Laboratory of Stanford University in Stanford, California. The obtained results were shown to be well reproduced by the calculations within the relativistic Fermi gas model of Ref.~\cite{Moniz:1969sr}, with two free parameters: the Fermi momentum and the average nucleon-separation energy.

Barreau \etal~\cite{Barreau:1983ht} at Saclay Linear Accelerator near Paris, France, performed extensive measurements of the inclusive cross section for electron scattering off carbon, using a range of energies and scattering angles. The goal of the study was to perform the Rosenbluth separation of the response functions including the $\Delta$-excitation region, to gain insight into both nucleonic degrees of freedom in the nucleus and meson-exchange currents, playing an important role in the dip region.

O'Connell \etal~\cite{O'Connell:1987ag} at Bates  Linear Accelerator Center in Cambridge, Massachusetts, extracted the $(e,e')$ cross sections for targets ranging from hydrogen to oxygen---including carbon---at two kinematic setups, and observed that the nuclear response per nucleon in the $\Delta$-excitation region is nearly the same for nuclei with mass numbers $A$ between 4 and 16.

Bagdasaryan \etal~\cite{Bagdasaryan:1988hp} at the Yerevan Electron Synchrotron in Armenia collected data for inclusive electron scattering off beryllium and carbon, covering both the QE and $\Delta$ peaks, and compared them with shell-model calculations.

Baran \etal~\cite{Baran:1988tw} at Stanford Linear Accelerator Center (SLAC) in Menlo Park, California,  performed a measurement of the inclusive cross section to analyze $\Delta$ excitation in carbon and iron nuclei, and separated the response functions. They observed that in the $\Delta$ region, the cross section scales with $A$, while in the dip region, it scales faster than $A$. Baran \etal{} also concluded that, within uncertainties, the measured cross section is completely transverse above the QE peak, and found indications of the importance of mechanisms involving removal of more than two nucleons from the nuclear ground state.

Sealock \etal~\cite{Sealock:1989nx} at SLAC studied inclusive excitation of $\Delta$ resonance for a range of targets, including carbon. They concluded that its peak position is---within uncertainties---independent of nuclear mass, but dependent on $Q^2$. In the dip region, Sealock \etal{} observed that the $A$ and $Q^2$ dependence indicates that the cross section receives a contribution that is specifically nuclear, and could be coming from proton-neutron pairs forming quasideuterons.

Day \etal~\cite{Day:1993md} at SLAC analyzed inclusive electron scattering covering a broad region of momentum transfer, systematically studying the dependence on nuclear mass number $A$, and performed extrapolation of the response functions to nuclear matter.

Gomez \etal~\cite{Gomez:1993ri} at SLAC performed a systematic study of the $A$ dependence of the EMC effect~\cite{Aubert:1983xm}, for targets ranging from deuteron to gold, including carbon. The results are reported both as the absolute cross sections and their ratios to deuteron, corrected for neutron excess.

Arrington \etal~\cite{Arrington:1995hs} at SLAC analyzed scaling in the Nachtmann variable $\xi$ of the inclusive scattering data collected for targets including deuteron and carbon at the kinematics corresponding to the Bjorken $x\simeq1$. Their results suggested a connection between QE and inelastic scattering, reminiscent of local duality in the nucleon~\cite{Bloom:1971ye}.

Arrington \etal~\cite{Arrington:1998hz,Arrington:1998ps} at Jefferson Lab (JLab) in Newport News, Virginia measured for a few targets (including deuteron and carbon) the inclusive cross section at large momentum transfers and showed that for sufficiently high $Q^2$, the collected data approach a scaling in $y$, $y$ being the minimum momentum of the struck nucleon along the direction of the momentum transfer~\cite{West:1974ua,Kawazoe:1975vv}. For the first time, it was also observed that the data exhibit a scaling behavior at very large negative $y$, where short-range correlations between nucleon pairs are expected to dominate the momentum distribution and final-state interactions.

Fomin \etal~\cite{Fomin:2008iq,Fomin:2010ei}  at JLab performed for several nuclei (including deuteron and carbon) an extensive measurement of the structure functions over a broad range of $\xi$ and $Q^2$ values, finding no evidence for extremely large contributions coming from short-range correlations, in agreement with high energy muon-scattering measurements, but in sharp contract to the findings of the CCFR E770 neutrino DIS experiment at Fermilab~\cite{Vakili:1999qt}.

Dai \etal~\cite{Dai:2018xhi} reported, from a measurement performed at JLab, the inclusive cross sections for carbon and titanium, extending from the QE peak to the region beyond the $\Delta$-excitation peak.

\begin{figure}
\centering
    \subfigure{\label{fig:comparisons_C_a_reducedDIS}}
    \subfigure{\label{fig:comparisons_C_b_reducedDIS}}
    \subfigure{\label{fig:comparisons_C_c_reducedDIS}}
    \includegraphics[width=0.90\columnwidth]{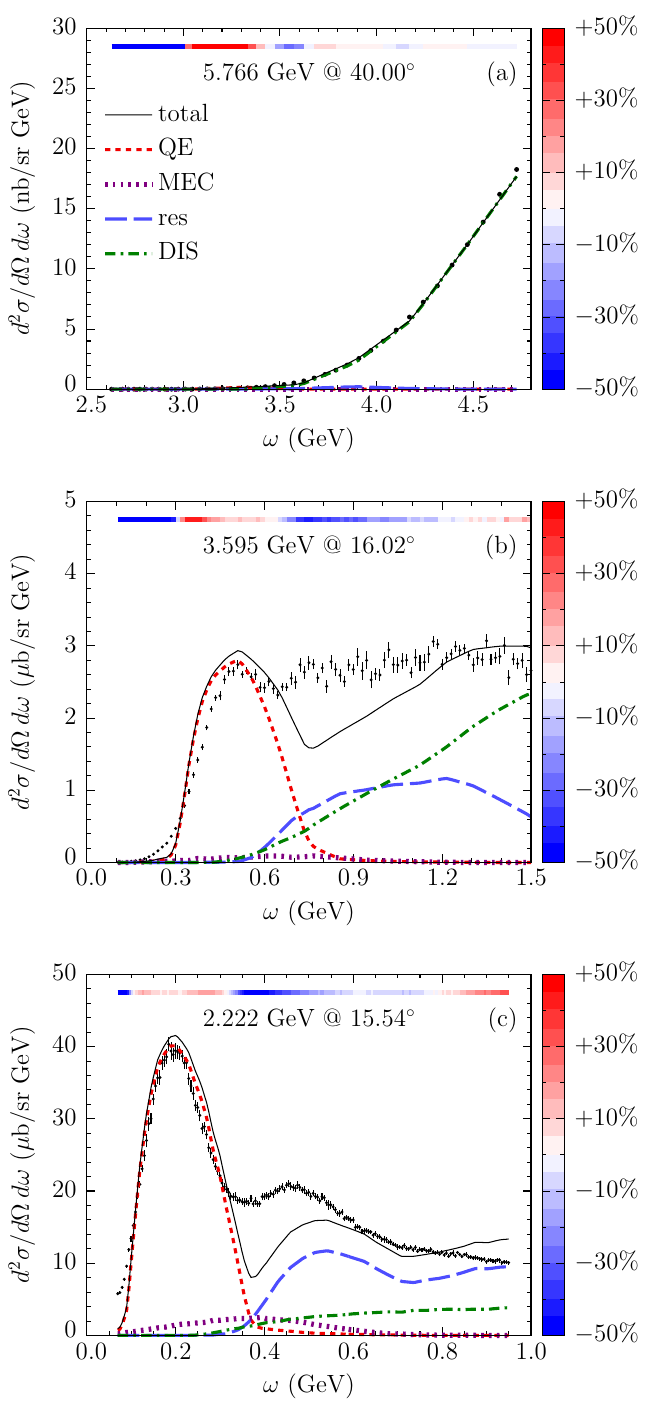}
\caption{\label{fig:comparisons_C_reducedDIS} Same as Fig.~\ref{fig:comparisons_C} but with the deep-inelastic contribution in \genie{} reduced by 28\%.}
\end{figure}

\begin{figure}
\centering
    \subfigure{\label{fig:qW_reducedDIS_a}}
    \subfigure{\label{fig:qW_reducedDIS_b}}
    \subfigure{\label{fig:qW_reducedDIS_c}}
    \includegraphics[width=0.85\columnwidth]{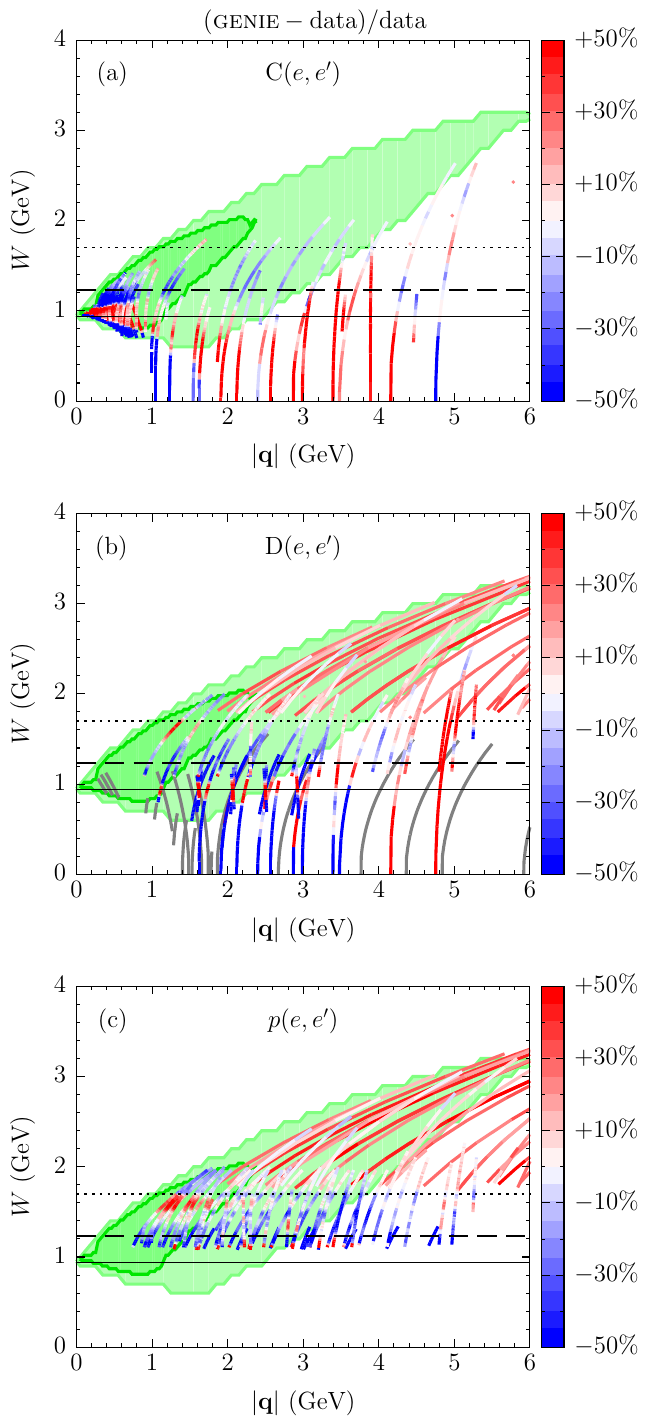}
\caption{\label{fig:qW_reducedDIS} Same as Fig.~\ref{fig:qW} but with the deep-inelastic contribution in \genie{} reduced by 28\%.}
\end{figure}

\subsection{Deuteron literature}
\label{Subsec:deuteron}
Whitlow \etal~\cite{Whitlow:1991uw} reanalyzed electron scattering data for proton and deuteron collected at SLAC by a series of measurements~\cite{Bodek:1973dy,Poucher:1973rg,Riordan:1974te,Stein:1975yy,Atwood:1976ys,Bodek:1979rx,Mestayer:1982ba,Arnold:1983mw,Gomez-thesis:1987,Dasu:1988ms}, applying improved procedures for radiative corrections. Making use of more precise knowledge of $R=\sigma_L/\sigma_T$~\cite{Whitlow:1990gk}, the ratio of the cross sections for the absorption of transverse and longitudinal photons, Whitlow \etal{} performed a global analysis and extracted the structure functions $F_2$ for proton and deuteron with significantly improved accuracy.

Sch{\"u}tz \etal~\cite{Schutz:1976he} at SLAC extracted the D$(e,e')$ cross section at the kinematics corresponding to high four-momentum transfer squared $Q^2$ and low energy transfer $\omega$, and observed scaling properties that were later associated with nucleon-nucleon short-range correlations~\cite{Frankfurt:1993sp}.

Parker \etal~\cite{Parker:1986} and Arnold \etal~\cite{Arnold:1988us} reported direct measurements---performed at Bates and SLAC, respectively---of the transverse response function by detecting electrons scattered off deuteron at $180^\circ$, to indicate the importance of mechanisms beyond the plane wave impulse approximation---such as those involving meson-exchange currents---in the dip region between the QE and $\Delta$-production peaks.

\begin{figure*}
\centering
    \includegraphics[width=0.90\textwidth]{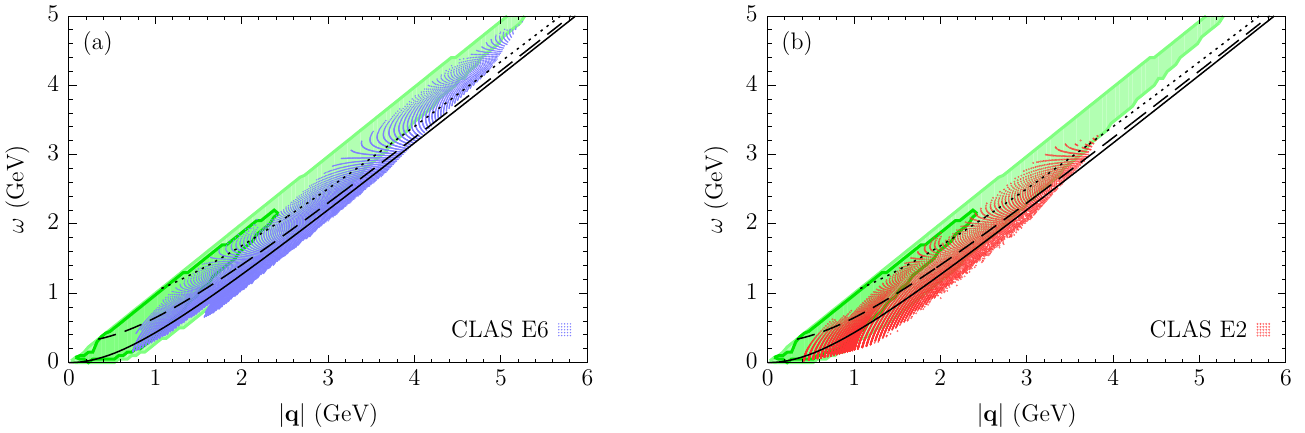}
\caption{\label{fig:prospects}Kinematic coverage of the CLAS measurements of electron scattering in the (a) deuteron experiment E6~\cite{Osipenko:2005gt} and (b) carbon experiment E2~\cite{Osipenko:2010sb} deduced from the data reported for the $F_2$ structure functions.}
\end{figure*}

Quinn \etal~\cite{Quinn:1988ua} at Bates collected the scattering data for both forward and backward angles, performed the Rosenbluth separation of the longitudinal and transverse response functions in the QE peak and the dip region, and compared them to different theoretical predictions.

Dytman \etal~\cite{Dytman:1988fi} at Bates obtained Rosenbluth separated response functions from the measured cross sections for deuteron and helium nuclei, and studied the dependence of the cross section in the QE peak and the dip region on momentum transfer $\n q$ and atomic number $A$.

Rock \etal~\cite{Rock:1991jy}  reported for high momentum transfers the cross section measured at SLAC over the region extending from the QE peak to the second resonance excitation, and analyzed its scaling properties.

Lung \etal~\cite{Lung-thesis:1992,Lung:1992bu} at SLAC performed measurements of the QE cross section at forward and backward scattering, separated the longitudinal and transverse response functions, and extracted the neutron electric and magnetic form factors.

Stuart \etal~\cite{Stuart:1996zs} reported measurements of inclusive electron-scattering cross sections for proton and deuteron in the $\Delta$ peak region, reanalyzing the data from Ref.~\cite{Lung-thesis:1992}.

Niculescu \etal~\cite{Niculescu:1999,Niculescu:2000tk} and Malace \etal~\cite{Malace:2006,Malace:2009kw} at JLab performed precision tests of quark-hadron duality in electron scattering off proton and deuteron, confirming that it holds both locally---for individual resonances---and globally---for the entire resonance region.

\subsection{Proton literature}\label{Subsec:proton}

Liang \etal~\cite{Liang:2003,Liang:2004tj} reported a detailed study of inclusive electron scattering in the resonance region, in which separated longitudinal and transverse response functions were obtained. This analysis found a substantial longitudinal component for resonances and observed quark-hadron
duality in the $F_1$ and $F_L$ structure functions independently.

\section{DIS reduction}
\label{appendix:DISReduction}
Let us discuss an example illustrating that tuning may mask underlying problems of Monte Carlo generators when the origin of the difference between the data and simulation is not understood. In this Appendix, we present \genie{} results with the DIS contribution reduced by 28\%.

Figure~\ref{fig:comparisons_C_reducedDIS} shows that this simplistic modification is sufficient to bring the predictions of \genie{} into much better agreement with the carbon data considered before in Fig.~\ref{fig:comparisons_C}. Rather consistent picture emerges also from the global comparison for carbon shown in Fig.~\ref{fig:qW_reducedDIS_a}. The reason for this behavior is that for this target, the data probing the DIS regime are scarce.

However, the global comparisons for deuteron and proton, presented in Figs.~\ref{fig:qW_reducedDIS_b} and~\ref{fig:qW_reducedDIS_c}, reveal that the problems in the DIS regime persist and cannot be resolved unless the functional dependence of the elementary cross sections is corrected.

\section{Other existing data}
\label{appendix:CLASdata}

In the process of preparation for the analysis presented in this paper, we have collected the total of 3,446 data points for carbon and 3,809 data points for deuteron, extracted by various experiments since the 1970s.

It is important to note that even more systematic comparisons would be possible, should the CLAS Collaboration reported the collected data in form of the cross sections. The studies of the $F_2$ structure functions of deuteron and carbon performed in the experiments E6~\cite{Osipenko:2005gt} and E2~\cite{Osipenko:2010sb}, respectively, turn out to cover a large swath of the kinematic region relevant for DUNE, as shown in Fig.~\ref{fig:prospects}.

With 12,120 data points published for deuteron~\cite{Osipenko:2005rh} and 9,934 for carbon~\cite{CLAS_database}, these experiments alone have the potential to become the most important source of our knowledge on how nuclear effects shape the cross sections for these nuclei. Such systematic information would be invaluable for estimating systematic uncertainties in neutrino-oscillation experiments.

\bibliographystyle{apsrev4-1}
\bibliography{bibliography}

\end{document}